\documentclass[aps,prx,superscriptaddress,twocolumn,nofootinbib,amsmath,amssymb,10pt,floatfix]{revtex4-2}

\pdfoutput=1
\usepackage{graphicx}
\usepackage{physics}
\usepackage{hyperref}
\usepackage{float}
\usepackage[table]{xcolor}
\usepackage{cleveref}
\usepackage{enumitem}
\usepackage{booktabs}
\usepackage{multirow}
\usepackage{etoolbox}
\robustify\bfseries
\usepackage{threeparttable}
\usepackage{siunitx}
\usepackage{placeins}
\usepackage{fontawesome}
\usepackage[defaultlines=3,all]{nowidow}

\usepackage{framed}
\usepackage[framed]{ntheorem}
\makeatletter
\newtheoremstyle{nonumberMyplain}
{\item[\theorem@headerfont\hskip\labelsep ##1\theorem@separator]\normalfont}%
{\item[\theorem@headerfont\hskip \labelsep ##1]{\theorem@headerfont (##3)}\theorem@separator\newline\normalfont}
\makeatother

\theoremstyle{plain}
\newtheorem{definition}{Definition}
\theoremstyle{nonumberMyplain}
\newframedtheorem{mapping}{Mapping:}
\newframedtheorem{synthesis}{Synthesis:}
\newframedtheorem{scheduling}{Scheduling:}
\newframedtheorem{architecture}{Architecture:}
\newframedtheorem{errors}{Errors}
\newtheorem{remark}{Remark}

\newcommand{\rwth}{Institute for Quantum Information, RWTH Aachen University, 52056 Aachen, Germany}
\newcommand{\fzj}{Peter Grünberg Institute, Theoretical Nanoelectronics, Forschungszentrum Jülich, 52425 Jülich, Germany}
\newcommand{\lmu}{Faculty of Physics, Ludwig-Maximilians-Univeristät, 80799 Munich, Germany}
\newcommand{\mpq}{Max-Planck-Institut für Quantenoptik, 85748 Garching, Germany}
\newcommand{\mcqst}{Munich Center for Quantum Science and Technology (MCQST), 80799 Munich, Germany}
\newcommand{\tum}{Chair for Design Automation, Technical University of Munich, 80333 Munich, Germany}
\newcommand{\scch}{Software Competence Center Hagenberg GmbH (SCCH), 4232 Hagenberg im Mühlkreis, Austria}

\newcommand{\CC}{C\nolinebreak\hspace{-.05em}\raisebox{.4ex}{\tiny\bf +}\nolinebreak\hspace{-.10em}\raisebox{.4ex}{\tiny\bf +}}
\newcommand*\rot{\rotatebox{90}}

\begin{document}

\title{Computational Capabilities and Compiler Development for Neutral Atom Quantum Processors --- Connecting Tool Developers and Hardware Experts}

\author{Ludwig Schmid} \email{ludwig.s.schmid@tum.de} \affiliation{\tum}
\author{David F. Locher} \email{d.locher@fz-juelich.de} \affiliation{\rwth} \affiliation{\fzj}
\author{Manuel Rispler} \email{rispler@physik.rwth-aachen.de} \affiliation{\rwth} \affiliation{\fzj}
\author{Sebastian Blatt} \email{sebastian.blatt@mpq.mpg.de} \affiliation{\lmu} \affiliation{\mpq} \affiliation{\mcqst}
\author{Johannes Zeiher} \email{johannes.zeiher@mpq.mpg.de} \affiliation{\lmu} \affiliation{\mpq} \affiliation{\mcqst}
\author{Markus Müller} \email{markus.mueller@fz-juelich.de} \affiliation{\rwth} \affiliation{\fzj}
\author{Robert Wille} \email{robert.wille@tum.de} \affiliation{\tum} \affiliation{\scch}

\begin{abstract}
\emph{Neutral Atom Quantum Computing} (NAQC) emerges as a promising hardware platform primarily due to its long coherence times and scalability.
Additionally, NAQC offers computational advantages encompassing potential long-range connectivity, native multi-qubit gate support, and the ability to physically rearrange qubits with high fidelity.
However, for the successful operation of a NAQC processor, one additionally requires new software tools to translate high-level algorithmic descriptions into a hardware executable representation, taking maximal advantage of the hardware capabilities.
Realizing new software tools requires a close connection between tool developers and hardware experts to ensure that the corresponding software tools obey the corresponding physical constraints.
This work aims to provide a basis to establish this connection by investigating the broad spectrum of capabilities intrinsic to the NAQC platform and its implications on the compilation process.
To this end, we first review the physical background of NAQC and derive how it affects the overall compilation process by formulating suitable constraints and figures of merit.
We then provide a summary of the compilation process and discuss currently available software tools in this overview.
Finally, we present selected case studies and employ the discussed figures of merit to evaluate the different capabilities of NAQC and compare them between two hardware setups.

\end{abstract}

\maketitle

\section{Introduction}\label{sec:introduction}
To achieve computational advantages with \emph{Quantum Computers} (QC), large-scale, high-fidelity qubit entanglement is required, posing a technologically challenging problem.
In recent years, qubit systems based on \emph{Neutral Atoms} (NA)~\cite{saffman_2016_quantum, morgado_2021_quantum} in combination with Rydberg interactions have established themselves as a promising candidate, due to their ability to perform high-fidelity long-range gates~\cite{graham_2022_multiqubit,levine_2019_parallel,evered_2023_highfidelity}, native multi-qubit gates~\cite{levine_2019_parallel,evered_2023_highfidelity,mueller_2009_mesoscopic,isenhower_2011_multibit}, and physical atom shuttling~\cite{bluvsteinLogicalQuantumProcessor2023,bluvsteinQuantumProcessorBased2022a}, combined with their scalability~\cite{ebadiQuantumOptimizationMaximum2022,barredo_2016_an_atom,endres_2016_atom}.

However, to fully harness these capabilities, it becomes essential to devise hardware-specific optimization techniques and software tools.
In particular, this includes \emph{compilation}, i.e.\ translating high-level algorithmic descriptions into a low-level representation of operations that can be executed on the hardware, obeying given physical constraints, and optimizing for specific figures of merit.
Manual optimization becomes infeasible as system sizes scale, necessitating automated processes and comprehensive toolkits to establish a complete compilation pipeline.
While a multitude of frameworks is available for other hardware platforms, such as \emph{superconducting} chips~\cite{Qiskit,willeMQTQMAPEfficient2023,cowtanQubitRoutingProblem2019,liTacklingQubitMapping2019,zulehnerEfficientMethodologyMapping2019,bergholmPennyLaneAutomaticDifferentiation2022,developersCirq2023,steigerProjectQOpenSource2018,tanOptimalLayoutSynthesis2020,sivarajahKetRetargetableCompiler2020}, or \emph{trapped ions}~\cite{sakiMuzzleShuttleEfficient2022,kreppel_2022_quantum,schmaleBackendCompilerPhases2022,maslovBasicCircuitCompilation2017,schoenbergerUsingBooleanSatisfiability2023}, the landscape of compiler tools tailored to NA-specific hardware constraints~\cite{liTimingAwareQubitMapping2023,patelGeyserCompilationFramework2022a,tanQubitMappingReconfigurable2022,brandhoferOptimalMappingNearTerm2021a,nottinghamDecomposingRoutingQuantum2023,bakerExploitingLongDistanceInteractions2021a,tanDepthOptimalAddressing2D2024,wangFPQACCompilationFramework2023,schmidHybridCircuitMapping2023} is still less developed.

Existing solutions for NAs often address specific compilation subproblems or make assumptions about a particular hardware configuration, failing to fully leverage the expansive capabilities of the NAQC platform.
To properly ensure that corresponding compilers and tools obey physical constraints and optimize for \mbox{hardware-specific} figures of merit, a close connection between tool developers and hardware experts is required.
The aim is to create valuable and high-quality compilation software that can leverage and take advantage of the full range of computational capabilities intrinsic to the NAQC platform.

This work aims to provide a basis to establish this connection by furnishing a comprehensive overview of software compilation for the NAQC platform and laying the groundwork for potential directions in compiler development geared explicitly toward adaptive, hardware-aware compilation strategies.
We discuss important figures of merit and employ them to evaluate the different capabilities of NAs, as well as compare them between two different hardware setups.

To achieve this goal, first, we establish a connection from physics to computer science by reviewing the physical background of NAs and translating the physical principles and processes to a hardware-aware but more abstract problem formulation suitable for tool developers.
The correspondingly resulting ``take home-messages'' for tool developers are then summarized 
in the form of optimization constraints and figures of merit in self-contained boxes, suitable résumés for individuals who are already familiar with NAs.
In this discussion, we particularly focus on the NA-specific capabilities of long-range connectivity, native multi-qubit gates, and the possibility of implementing MOVE operations on the qubits, using shuttling.

Secondly, we discuss how these new capabilities impact the compilation task, give a comprehensive overview of the full range of the compilation possibilities, and contextualize currently available software within this overview.
This discussion gives a possibility to structure the current progress of compilation development and aids potential tool developers in identifying unsolved subproblems and automation tasks.

Finally, we present selected case studies and error analysis to provide an overview of the current state of the art of hardware-aware compilation for NAQC.
This provides insights for tool developers regarding suitable figures of merit and optimization metrics depending on available hardware properties. 
Concurrently, this gives hardware experts the means to estimate the impact of their hardware configuration on the final compilation output and could potentially aid in devising forthcoming hardware arrangements, prioritizing hardware attributes that yield the most encouraging outcomes.

Based on these three main contributions, we provide the basis for a successful connection between tool developers and hardware experts for physical realization, which is necessary for the future development of hardware-aware compilation tools, taking full advantage of the broad spectrum of capabilities of the NAQC platform.

The remainder of this work is structured as follows:
In \Cref{sec:compilation}, we discuss the specific compilation aspects addressed in this work.
We define key terms and compilation steps, such as synthesis, mapping, and routing, at an abstract and hardware-independent level.
\noindent In \Cref{sec:neutral-atom-quantum}, we focus on the NAQC platform.
Here, we explore its distinctive capabilities and characteristics relevant to compilation.
Each subsection covers a particular capability and discusses its underlying physical principles and mechanisms.
This is followed by an abstract formulation tailored to the computer science community, summarized in self-contained boxes referring to the three previously introduced compilation steps.
\noindent In \Cref{sec:compilation-for-na}, we investigate how these novel capabilities influence the compilation process, outlining a potential overview and illustrating the overall compilation problem in an overview figure.
Referring to this framework, \Cref{sec:rel-work-software} reviews existing software and discusses what capabilities they focus on,  giving a  comprehensive overview of currently available software for NAQC.
\noindent \Cref{sec:numer-eval} offers selected case studies that showcase how the NA platform's various capabilities and hardware parameters influence the computation results.

\section{Compilation}
\label{sec:compilation}

In its broadest sense, the term ``compilation'' refers to translating a high-level, abstract description of a quantum algorithm into a lower-level representation of operations suitable for hardware execution.
This is typically achieved by employing multiple layers of software, each designed to address specific subroutines.
The collective arrangement of these layers is commonly known as a \emph{compilation tool-chain}~\cite{chongProgrammingLanguagesCompiler2017}.
Given the diverse range of tasks involved and the intersection between computer science- and physics-related terminology, various terms, including the term ``compilation'' itself, can vary depending on the context.
To avoid ambiguity, we provide a brief review and definitions of different subroutines considered in this work.

\subsection{Compilation Subroutines}
\label{sec:comp-soubr}
The subroutines can be divided into three major categories, depending on their abstraction level and their specificity to a particular hardware configuration.

\begin{itemize}
  \item{\emph{Platform-independent compilation}}\\
        High-level optimization techniques can be applied irrespective of the underlying platform.
        These techniques include for example loop unrolling and function inlining~\cite{javadiabhariScaffCCFrameworkCompilation2014}, to substitute or simplify e.g.,~for-loops and function calls on a high level similar to classical compilers. 
        Another possibility are optimization techniques on the gate level such as gate commutation rules.
  \item{\emph{Platform-dependent compilation}}\\
        These subroutines account for a given hardware platform's specific capabilities and constraints, such as superconducting chips, trapped ions, NAs, or photonic quantum computing.
        The output of these subroutines comprises a sequence of platform-specific instructions, which is still agnostic of the physical hardware setup.
  \item{\emph{Hardware-dependent compilation}}\\
        This translation layer is tailored to each hardware configuration, converting abstract quantum operations into hardware-specific instructions that can be directly executed on the quantum processing unit.
        It is occasionally referred to as \emph{firmware} to underscore its close proximity to the underlying hardware.
\end{itemize}

This work focuses on addressing the subroutines of platform-dependent compilation, particularly the NAQC platform.
Specifically, we assume that all hardware- and platform-independent optimizations have already been conducted, and our objective is to generate a collection of hardware-oriented operations without delving into discussions related to direct hardware control through electric signals or pulse-level intricacies.
This leads to three primary objectives.
\begin{enumerate}
  \item\emph{Synthesis} entails decomposing abstract gates into operations compatible with the provided platform and is therefore also often referred to as \emph{decomposition}.

  \item\emph{Mapping} involves spatially arranging the gates by assigning circuit qubits to corresponding hardware qubits and inserting SWAP or MOVE operations to fulfill connectivity constraints.
  \item \emph{Scheduling} corresponds to the temporal arrangement of the gates to satisfy the dependencies and consider the parallelism constraints inherent to the platform or the hardware.
\end{enumerate}

\begin{figure*}[ht]
\centering
\includegraphics[width=0.95\textwidth]{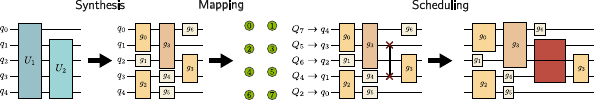}
\caption{\textbf{Illustration of the three steps for platform-dependent compilation.} In the \textbf{synthesis} step, general operations and unitaries are decomposed into the native gate set $\Sigma_{\mathrm{native}}$.
  During the \textbf{mapping} step, the circuit qubits $q_{i}$ are assigned to physical hardware qubits $Q_{i}$, and necessary SWAP or MOVE operations are introduced to satisfy connectivity constraints.
  Finally, in the \textbf{scheduling} step, gate times and restrictions on parallelism are considered.
  In practice, these steps are often performed simultaneously as a single step rather than sequentially.}
\label{fig:compilation_steps}
\end{figure*}

In the following, we present concise and abstract definitions of these three steps tailored to the context of this work.
An illustration is shown in \Cref{fig:compilation_steps}.

\subsubsection{Synthesis}
\label{sec:synthesis}
Within the framework of the \emph{quantum circuit model}, every non-dissipative quantum computation $U \in \mathbb{C}^{2^n\times 2^n}$ can be expressed as a finite sequence of quantum operations $g \in \mathbb{C}^{2^m \times 2^m}$ known as quantum gates, denoted by reversible unitary transformations.
Here,  $n$ ($m$) denotes the number of qubits upon which the circuit (gate) acts.
According to the Solovay-Kitaev theorem~\cite{kitaevQuantumComputationsAlgorithms1997}, given a \emph{universal gate set} $\Sigma^{\mathrm{univ}}$ with a discrete number of gates, an approximate decomposition up to an arbitrarily small error can always be obtained.
Furthermore, this can be done efficiently in terms of the number of gates.
In contrast, the \emph{native gate} set $\Sigma^{\mathrm{native}}$ characterizes the feasible operations that can be performed on the quantum state using a specific hardware platform or setup.
For universal computing, it is necessary that $\Sigma^{\mathrm{native}}$ is also a universal gate set.

\begin{definition}[Synthesis]
  Given a quantum computation $U \in \mathbb{C}^{2^n\times2^n}$ and the native platform gate set $\Sigma^{\mathrm{native}}$, \emph{synthesis} is the task to find a gate sequence
  \begin{equation*}
    \label{eq:synthesis}
    \tilde{U} = g_{N-1} \circ \dots  \circ g_{0}
  \end{equation*}
  with all $g_{0}, \dots g_{N} \in \Sigma^{\mathrm{native}}$ and $U = \tilde{U}$ up to some small error.
\end{definition}

\subsubsection{Mapping}
\label{sec:-mapping}
Each gate operates on a subset of the \emph{circuit qubits} denoted as $\mathbf{Q} = \{q_{0}, ..., q_{n-1}\}$.
For instance, $g_{0}(q_{0},q_{1})$ indicates that the first gate of the computation acts on the circuit qubits $q_{0}$ and $q_{1}$.
These gates must be implemented using the available physical \emph{hardware qubits} represented by $\mathbf{P} = \{Q_{0}, ..., Q_{n-1}\}$. Without loss of generality, we assume that the number of circuit and physical qubits is the same.
A common challenge encountered in current hardware platforms is limited connectivity, which is described by a \emph{coupling graph} $G = ({\mathbf{P}, \mathbf{E}})$.
In this graph, the nodes correspond to the physical qubits, while the edges $\mathbf{E}$ indicate the qubits capable of interacting with each other.

To execute a gate $g(Q_{i}, Q_{j})$ with $(Q_{i}, Q_{j}) \notin \mathbf{E}$ (that is, the qubits are not directly connected), it is necessary to adjust the coupling graph to establish the required connectivity.
In most platforms, connectivity is closely related to the physical locations of the qubits.
Therefore, two commonly employed techniques are SWAP and MOVE operations.
The $\text{SWAP}(Q_{i}, Q_{j})$ operation exchanges the positions of the qubits $Q_{i}$ and $Q_{j}$, resulting in a modification of their labels in the coupling graph.
On the other hand, the $\text{MOVE}(Q_{i})$ operation relocates qubit $Q_{i}$ to a different position, consequently reassigning the associated edges.

In superconducting (SC) hardware, these operations are typically performed at the virtual level, acting on $\mathbf{Q}$, and require, for example, three controlled-NOT (CX) gates to implement a SWAP operation.
On the contrary, for other platforms, such as trapped ions or NAs, it may be possible to physically move and swap the corresponding atoms or ions, directly affecting the hardware qubits $\mathbf{P}$.

\begin{definition}[Mapping]

  Given a quantum circuit ${U= g_{N-1} \circ \dots \circ g_{0}}$ on circuit qubits $\mathbf{Q}$ and a hardware configuration with physical qubits $\mathbf{P}$ and coupling map $G(\mathbf{P}, \mathbf{E})$.
  The task of \emph{mapping} is to find a bijective function $f: \mathbf{Q} \rightarrow \mathbf{P}$ and an insertion of MOVE and SWAP operations such as
  \begin{equation*}
    \label{eq:moveswap}
    U = \text{...} \circ \text{MOVE}(q_{i}) \circ \text{SWAP}(q_{j},q_{k}) \circ g(q_{i},q_{j}) \circ \text{...}
  \end{equation*}
  such that for each gate $g(q_{i},q_{j})$ all inter-qubit connections are fulfilled, i.e.~$(f(q_{i}),f(q_{j})) \in \mathbf{E}$.
\end{definition}

It should be noted, that this graph-based approach to mapping can only represent a first approximation. 
In general and, in particular, for multi-qubit gates, additional constraints such as gate direction and the geometric arrangement of the qubits can impose additional constraints.

\subsubsection{Scheduling}
\label{sec:scheduling}
While mapping primarily concerns the spatial arrangement of gates on qubits to satisfy connectivity constraints, we must also consider the temporal positioning of the gates.
Subject to commutation rules, gates acting on the same qubit must be executed in a specific order.
This aspect can be abstracted by transforming the gate sequence ${U = g_{N-1} \circ \dots \circ g_{0}}$ into a \emph{Directed Acyclic Graph} (DAG) denoted $D$.
In this DAG, the nodes correspond to quantum gates, while the incoming and outgoing edges correspond to the qubits on which the gates operate.
The direction of the edges reflects the sequential execution order of the gates.

When two gates act on disjoint sets of qubits, lacking any common path in the DAG, they can generally be executed in parallel.
However, the execution of gates in parallel may face additional constraints imposed by hardware limitations, such as the availability of control channels for qubit control, or platform-specific restrictions on gate operations, such as cross-talk effects.

\begin{definition}[Scheduling]

Given a quantum circuit $U$ and its corresponding DAG representation $D$, the objective of \emph{scheduling} is to determine the optimal timing for the gates to be executed while preserving the integrity of the DAG up to commutation rules.

\end{definition}

Due to decoherence, the primary goal during the scheduling phase is typically to maximize parallelism, thereby minimizing the overall execution time of the circuit.

In the next section, we delve into a comprehensive study of the NAQC platform, exploring its unique computational capabilities compared to other platforms.
In particular, we analyze how the physical principles and processes of the NAs can be formulated on an abstract level regarding optimization constraints and figures of merit, focusing on the three compilation steps of synthesis, mapping, and scheduling.

\section{The Neutral Atom Quantum Computing Platform}\label{sec:neutral-atom-quantum}

Building on continuous efforts in using NAs for atomic clocks and analog quantum simulations, several recent experimental and theoretical breakthroughs have allowed this platform to evolve into a promising candidate for scalable digital quantum computing \cite{graham_2022_multiqubit, saffman_2010_quantum, saffman_2016_quantum, adams_2020_rydberg, henriet_2020_quantum, morgado_2021_quantum, wu_2021_a_concise, levine_2019_parallel,levine_2018_highfidelity}.
This section provides an overview of the fundamental physical principles essential for employing NAs in quantum computing and the resulting additional capabilities.
Based on that, we then translate these physical principles and processes to a hardware-aware but more abstract problem formulation, establishing a connection between physics and computer science and, hence, hardware experts and tool developers. This results in ``take-home messages'' for tool developers, which are provided in terms of dedicated boxes covering optimization constraints as well as figures of merit and entailing everything relevant for compiler development.
Individuals with prior knowledge of NAs and/or compilation for quantum hardware may only check these boxes or bypass the corresponding sections completely.
In the following, we first show that the NAQC is a suitable candidate for universal quantum computation.
We discuss in depth the NA capabilities and error sources, each entailed by one of the aforementioned summary boxes focusing on the three hardware-dependent compilation steps of synthesis, mapping, and routing, extended by hardware and error discussions.

\subsection{Platform Requirements}\label{sec:platform-requirements}
For the NAQC to be suitable as a quantum computing platform, one requires multiple properties, which are commonly known as DiVincenzo's criteria \cite{divincenzo_2000_the_physical_implementation}, namely:
\begin{enumerate}
    \item A scalable physical system of qubits
    \item The ability to initialize the state of the qubits to a simple fiducial state
    \item A universal set of quantum gates
    \item Long relevant coherence times, much longer than the gate operation time
    \item A qubit-specific measurement capability.
\end{enumerate}
The following outlines that all those requirements are fulfilled for NAs trapped in optical tweezer arrays or optical lattices.

\noindent
\emph{1. System and qubits:}
NAs can be stored in optical dipole traps~\cite{grimm_2000_optical}, which are typically formed by far-detuned lasers interacting with the atomic dipole moment.
Examples include optical lattices, which are stationary patterns of light that emerge from the interference of multiple laser beams \cite{jessen_1996_optical}, or optical tweezer arrays, which consist of many individual tightly focused and individually controllable laser beams arranged in one, two, or three dimensions \cite{kaufman_2021_quantum}.
In state-of-the-art optical tweezer arrays, atoms are loaded stochastically and can then be dynamically rearranged in desired configurations \cite{barredo_2016_an_atom, endres_2016_atom}, with the possibility of dynamically reloading atoms from a reservoir to compensate for occasional loss of atoms \cite{gygerContinuousOperationLargescale2024,norciaIterativeAssembly1712024}.
It is possible to laser-cool the atoms within the traps to their motional ground states \cite{kaufman_2012_cooling}.
Commonly used atom species in such experiments are alkali atoms, e.g.~Rb, Cs, or alkaline-earth-like atoms, such as Sr or Yb.
Specific internal electronic states of the atoms serve as the qubit states $\ket{0}$ and $\ket{1}$.
Both states are typically encoded in two long-lived states of the atom, such as two hyperfine states in alkali atoms~\cite{levine_2019_parallel, graham_2019_rydberg}, a nuclear spin~\cite{barnes_2022_assembly, ma_2022_universal,jenkins_2022_ytterbium} ,low-energy singlet and triplet states~\cite{young_2020_halfminute, schine_2022_longlived}, fine-strucure~\cite{pucherFineStructureQubitEncoded2024,unnikrishnanCoherentControlFineStructure2024}, or circular Rydberg states~\cite{holzlLongLivedCircularRydberg2024} in alkaline-earth (like) atoms.
Due to their different properties, also dual-species atom arrays have been proposed and demonstrated~\cite{anandDualspeciesRydbergArray2024}.

\noindent
\emph{2. Initialization, and 3. Quantum gates:}
Transitions between electronic states can be precisely controlled by applying laser pulses to the atoms.
Such pulses are used to initialize qubits in well-defined initial states, e.g.~$\ket{0}$, and to perform arbitrary single-qubit gates.
Universal quantum computing also demands controlled two-qubit gates.
To achieve this, atoms can be temporarily excited to Rydberg states, which are electronic states with very large principal quantum numbers.
Atoms in Rydberg states exhibit large polarizabilities, and as a consequence, two Rydberg atoms interact via dipole-dipole interactions~\cite{gallagher_1994_rydberg}.
By coupling one of the qubit states, e.g.~$\ket{1}$, to a Rydberg state $\ket{r}$, one obtains an effective interaction between two atoms in the state $\ket{1}$ \cite{jaksch_2000_fast,saffman_2010_quantum}. Depending on the exact settings, interaction characteristics can vary from dipolar to van der Waals~\cite{morgado_2021_quantum}.
This mechanism can be used to engineer two-qubit or multi-qubit gates.

\noindent
\emph{4. Coherence times:}
The two primary error sources of trapped atoms during idling are dephasing and amplitude damping ($\ket{1} \mapsto \ket{0}$).
These processes cause the off-diagonal terms in the density matrix $\rho$ of a qubit to decay on a time scale $T_2^*$, and the matrix element $\rho_{11}$ to decay on a time scale $T_1$.
Both time scales can reach up to the order of several seconds for NAs in optical tweezers~\cite{young_2020_halfminute,evered_2023_highfidelity} and are characteristic for the specific qubit implementation, e.g.~magnetically insensitive clock states with very long lifetimes.
Controlled phase gates can be performed in timescales of the order of $\SI{100}{ns} - \SI{500}{ns}$~\cite{madjarov_2020_highfidelity}, depending on the physical details, which is orders of magnitude faster than the decoherence time of such systems.

\noindent
\emph{5. Measurements:}
Finally, the qubit states of the atoms can be measured by performing fluorescence imaging, that is, driving transitions between one of the qubit levels and an auxiliary electronic level with a laser and imaging the emitted photons with a camera.

\subsection{Computational Capabilities}\label{sec:comp-capabilities}

\begin{figure*}[ht]
\centering
\includegraphics[width=\textwidth]{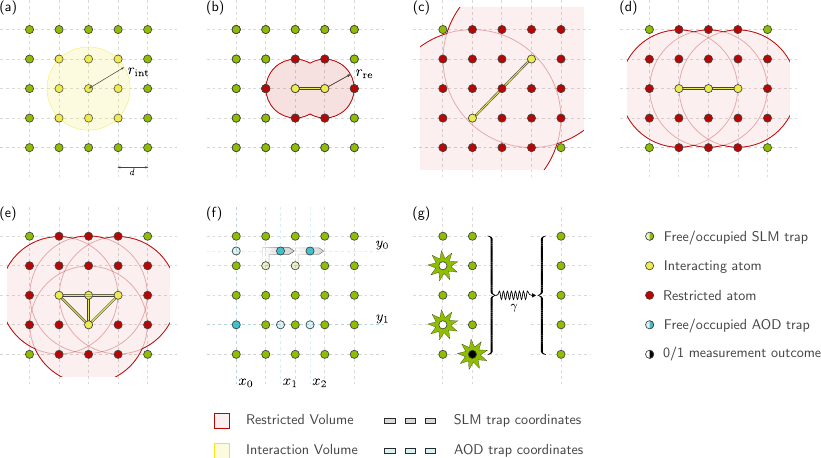}
\caption{
  \textbf{Capabilities of the NAQC platform.} In this setup, atoms are arranged on a regular grid of Spatial Light Modulator (SLM) traps, with a fixed distance denoted as $d$.
  \textbf{(a)} Rydberg blockade interaction: Within a specific interaction zone of radius $r_{\mathrm{int}} = \sqrt{2}d$, depicted in yellow, interacting gates can be performed to all qubits within this range.
  \textbf{(b)} Two-qubit gate: A gate can be applied between neighboring qubits but restricts the simultaneous execution of other entangling gates on nearby atoms.
  The restriction volume, represented by a red sphere of radius $r_{\mathrm{re}} = d$, indicates this restriction.
  The interaction radius $r_{\mathrm{int}}=d$ is not explicitly shown to simplify the illustration. \textbf{(c)} Long-Range interactions: For gates with larger interaction radii, the restriction zones ($r_{\mathrm{re}} = 3d \geq r_{\mathrm{int}} = 2\sqrt{2}d$) also expand, resulting in more restricted atoms.
  \textbf{(d)} CCZ gate with a line arrangement of the qubits. 
  According to \cite{levine_2019_parallel} it is sufficient if the central atom interacts with both the outer qubits, resulting in a minimal interaction radius of $r_\mathrm{int} = d$. 
  The restriction radius is illustrated examplatory as $r_\mathrm{re}= 2d$.
  \textbf{(e)} CCCZ gate: In this case, we require that all four gate qubits must be within the interaction radius $r_{\mathrm{int}} = r_{\mathrm{re}}=2d$ of every other qubit, according to \Cref{eq:multi-qubit-mapping}.
  \textbf{(f)} Shuttling operation: Dynamic Acousto-Optic Deflector (AOD) traps (blue) enable the movement of atoms within the same column ($x$) or row coordinate ($y$) simultaneously.
  The procedure addresses certain constraints discussed in \Cref{sec:atom-shuttling}.
  \textbf{(g)} Additional NA capabilities, useful for future fault-tolerant computations.
  For example, mid-circuit measurements and possible inter-photonic connections.
}\label{fig:capabilities}
\end{figure*}

In addition to fulfilling DiVincenzo's criteria, NAs offer a broad spectrum of capabilities that can be employed to perform quantum computations.
In the following, these capabilities are discussed in more detail.
This is done by first discussing the corresponding capabilities' basic physical principles and processes, followed by self-contained boxes, summarizing the resulting constraints and optimization figures of merit on a more abstract level.
Therefore, tool developers can refer only to the boxes for the development of NA-specific compilers.

\subsubsection{Single-qubit Gates}
\label{sec:single-qubit-gates}

Single-qubit gates are realized with lasers that drive Rabi oscillations between the qubit states $\ket{0}$ and $\ket{1}$.
The respective Hamiltonian reads
\begin{equation} \label{eq:single_qubit_hamiltonian}
    \frac{H_1(t)}{\hbar} = \frac{\Omega(t)}{2} \ketbra{0}{1} + \frac{\Omega^{*}(t)}{2} \ketbra{1}{0} - \Delta(t) \ketbra{1}{1} ,
\end{equation}
where $\Omega(t)$ and $\Delta(t)$ are the effective Rabi frequency and detuning, respectively.
Depending on the precise qubit encoding, the transitions might be either single- or two-photon transitions.
In the latter case, the Hamiltonian in \Cref{eq:single_qubit_hamiltonian} is obtained after adiabatic elimination of an intermediate level.
Depending on the experimental setup, some laser beams might only be available as global beams~\cite{graham_2022_multiqubit}, simultaneously illuminating many atoms.
However, single-qubit addressing can also be realized using beams focused on individual atoms~\cite{wangIndividualatomControlArray2023,tanDepthOptimalAddressing2D2024}.

\subsubsection{Two-qubit Gates} \label{sec:2qubitgates}

To execute two-qubit gates, typically, one of the qubit states, e.g.~$\ket{1}$, is coupled to a \emph{Rydberg state} $\ket{r}$ using a laser with Rabi frequency $\Omega_{\mathrm{r}}$ and detuning $\Delta_{\mathrm{r}}$.
Two atoms in the same Rydberg state and separated sufficiently far from each other interact via the van der Waals interaction
\begin{equation} \label{eq:van_der_Waals}
    V_{\mathrm{vdW}}(d) = \frac{C_6}{d^6} ,
\end{equation}
where $C_6$ is a state-specific constant and $d$ the inter-qubit distance.
The Hamiltonian describing two atoms illuminated by a global Rydberg laser, therefore, reads
\begin{equation}
    \begin{aligned}
    \frac{H_2(t)}{\hbar} = \sum_{i=1}^{2} &  \hspace{-2pt} \left( \frac{\Omega_{\mathrm{r}}(t)}{2} \ketbra{1}{r}_{i} + \frac{\Omega^{*}_{\mathrm{r}}(t)}{2} \ketbra{r}{1}_{i} - \Delta_{\mathrm{r}}(t) \ketbra{r}{r}_{i} \right) \\ 
    & + \frac{V_{\mathrm{vdW}}}{\hbar} \ketbra{r}{r}_1 \otimes \ketbra{r}{r}_2 .
    \end{aligned}
\end{equation}
The van der Waals interaction imposes an energy penalty to promote two nearby atoms together to a Rydberg state.
Consequently, laser excitation of one atom to a Rydberg state is impossible if another Rydberg atom is close by.
This so-called \emph{Rydberg blockade} can be utilized to engineer state-dependent interactions and, thus, two-qubit gates between two atoms.
The first proposal by~\citet{jaksch_2000_fast} requires a sequence of locally addressed laser pulses and was realized for the first time in 2010 with neutral Rb atoms \cite{isenhower_2010_demonstration, wilk_2010_entanglement}.
More recent two-qubit gate sequences require only global addressing of both involved qubits and are also faster than the original scheme~\cite{levine_2019_parallel}.
The qubits must be located sufficiently close to each other so that the interaction is strong enough for the Rydberg blockade to hold.
From the condition $\hbar \Omega_{\mathrm{r}} \simeq V_{\mathrm{vdW}}(r_{\mathrm{b}})$ follows a blockade radius
\begin{equation}
    r_{\mathrm{b}} \simeq \left( \frac{C_6}{\hbar \Omega_{\mathrm{r}}}\right)^{1/6} .
\end{equation}
This length scale gives an estimate of how far two atoms participating in a two-qubit gate can be separated at most.
Since the blockade radius can be much larger than the distance between neighboring atoms, two-qubit gates can be applied to pairs of atoms that are not nearest neighbors, resulting in higher inter-qubit connectivity.
\newpage

\begin{synthesis}[One- and two-qubit gates]
  Assuming local addressability of all qubits, this results in a native gate set $\Sigma^{\mathrm{univ}}= \{R, CZ\}$ containing arbitrary single-qubit rotations $R$ and the two-qubit CZ gate to synthesize a given algorithm.

  Nevertheless, synthesis for NAs is hard, as the exact synthesis steps required depend heavily on the chosen atom species and qubit encoding.
  Furthermore, one has to distinguish between laser pulses available as global or locally addressable beams.
  A first approach, leveraging SC synthesis tools with additional post-processing was demonstrated by~\citet{nottinghamDecomposingRoutingQuantum2023}.
\end{synthesis}

\begin{mapping}
  Two-qubit gates can be executed between any pair of qubits sufficiently close for the blockade to take effect.
  This constraint can be described by a constant \emph{interaction radius} denoted as $r_{\mathrm{int}}$, where a gate $g(Q_{i},Q_{j})$ can be performed if
  \begin{equation}
    \label{eq:interaction_radius}
    d(Q_{i}, Q_{j}) \leq r_{\mathrm{int}},
  \end{equation}
  where $d$ is the Cartesian distance.
  This is illustrated in \Cref{fig:capabilities} (a).
  
  We can define a coupling graph $\mathbf{C} = (V,E)$, with
  \begin{equation}
    \label{eq:coupling_graph}
      E = \{(Q_{i},Q_{j}) | d(Q_{i},Q_{j}) \leq r_{\mathrm{int}}\}.
  \end{equation}
  With this coupling graph, common mapper tools of the SC platform\cite{Qiskit,willeMQTQMAPEfficient2023,bergholmPennyLaneAutomaticDifferentiation2022,developersCirq2023,liTacklingQubitMapping2019} can be partially employed for the NAQC platform.
\end{mapping}

However, during the two-qubit gate sequence, both involved qubits are promoted to the Rydberg state and, therefore, could interact with other nearby atoms in Rydberg states, resulting in unwanted detrimental cross-talk.
While the strong Rydberg interaction is indispensable for achieving high-fidelity two-qubit gates, it also imposes limitations on other atoms in close proximity, preventing them from simultaneously performing similar gates.

\begin{scheduling}
  \label{scheduling:blocking}
  To establish the constraint on simultaneous gate executions, we use the concept of a \emph{restriction radius} denoted as $r_{\mathrm{re}}$.
  This parameter represents the minimum distance required between two atoms in the Rydberg state to prevent undesirable cross-talk.
  It is worth noting that, in general, $r_{\mathrm{re}} \geq r_{\mathrm{int}}$, as cross-talk may arise even at distances where a gate interaction may not yet be feasible.

  Thus, for two gates $g(Q_{i}, Q_{j})$ and $g'(Q_{a}, Q_{b})$ to be executed in parallel, the following condition must be satisfied:
  \begin{equation}
    \label{eq:blocking}
    \hspace{-5pt} d(Q_{i},Q_{a}), d(Q_{i},Q_{b}),d(Q_{j},Q_{a}),d(Q_{j},Q_{b}) > r_{\mathrm{re}}.
  \end{equation}
  If the distances between qubits involved in gates are too small, gates must be executed in sequential order, possibly increasing the execution time of the circuit.
  This limitation is commonly known as \emph{blocking}; however, to avoid any potential confusion with the concept of Rydberg blockade, we use the term \emph{restriction}.
  We refer to the region surrounding a gate, affected by this phenomenon, as the \emph{restriction volume}.
  
  A simple visual example with different radii is illustrated in \Cref{fig:capabilities} (b) and (c), where the restriction volume and the corresponding restricted atoms are colored in red.
\end{scheduling}

The interdependence of $r_{\mathrm{int}}$ and $r_{\mathrm{re}}$ on the strength of the Rydberg blockade $r_{\mathrm{b}}$ allows us to establish a relationship between them as $r_{\mathrm{re}} = k \cdot r_{\mathrm{int}}$, where $k \geq 1$ is referred to as \emph{blocking factor}.
This gives rise to an interesting trade-off between higher connectivity achieved with a larger $r_{\mathrm{int}}$, which, in turn, leads to larger restriction volumes and consequently reduces the number of parallel gate executions.
This phenomenon has been investigated by \citet{bakerExploitingLongDistanceInteractions2021a} and is further discussed in greater detail with additional case studies and error analysis in \Cref{sec:numer-eval}.

\subsubsection{Multi-qubit Gates}\label{sec:multi-qubit-gates}

In NA quantum processors, it is also possible to apply native multi-qubit gates to sets of atoms.

Analogous to the case of two-qubit gates, the atoms must be located within the blockade radius of each other such that the atoms interact when excited to Rydberg states.
There exist multiple theoretical proposals to implement multi-qubit gates such as $\mathrm{C}_{k}\mathrm{Z}$ and $\mathrm{C}\mathrm{Z}_{k}$ gates, up to single-qubit rotations, and also more exotic versions \cite{mueller_2009_mesoscopic, isenhower_2011_multibit, dlaska_2022_quantum, evered_2023_highfidelity}.
Parallel applications of Toffoli (CCZ) gates to multiple sets of atoms have already been demonstrated in experiments with trapped Rb atoms \cite{levine_2019_parallel, evered_2023_highfidelity}.

\begin{synthesis}
  This expands the possible native gate set to $\Sigma^{\mathrm{univ}}= \{R, \mathrm{C}_{k}\mathrm{Z},\mathrm{C}\mathrm{Z}_{k}\}$ and, therefore, gives much more possibilities to synthesize a given circuit.
  First, multi-qubit gates within an algorithm do not have to be decomposed to a more basic gate set, possibly increasing gate or decoherence errors.
  
  Furthermore, native multi-qubit gates are of great interest for the implementation of classical reversible circuits, using Toffoli and multiple controlled gates~\cite{shendeReversibleLogicCircuit2002,adarshSyReCSynthesizerMQT2022,willeRevLibOnlineResource2008,willeSyReCProgrammingLanguage2010}.
  Those classical circuits can be used for algorithmic subtasks or the synthesis of oracles used in quantum algorithms. 
  In this context, oracles refer to black-box functions performing the required operation.
  Although often used during the construction of quantum algorithms, the task of finding an adequate realization is non-trivial in general.\\
  Secondly, instead of decomposing unitaries to one- and two-qubit gates only, also other gate combinations such as Hadamard plus Toffoli~\cite{amyImprovedSynthesisToffoliHadamard2023,aharonovSimpleProofThat2003} may be used, resulting in alternative decompositions.
\end{synthesis}

\begin{mapping}
  The mapping constraints for multi-qubit gates depend on the specific implementation and the corresponding laser pulses. To simplify the discussion we assume the worst case, where each qubit needs to interact with each other qubit within the gate.
  Therefore, to implement a multi-qubit gate $g(Q)$ acting on the set of qubits \mbox{$Q = \{Q_{i}, \dots, Q_{j}\}$}, all qubits have to be within the interaction radius of each other:
  \begin{equation}
    \label{eq:multi-qubit-mapping}
    d(Q_{i},Q_{j}) \leq r_{\mathrm{int}} \, \forall i,j \in Q .
  \end{equation}
  Furthermore, this implies that $r_{\mathrm{int}}$ places a constraint on the size and configuration of multi-qubit gates.
  For instance, a Toffoli gate with its three qubits aligned in a straight line necessitates $r_{\mathrm{int}} \geq 2d$ as in \Cref{fig:capabilities} (d), while a possible perpendicular arrangement could be achieved with $r_{\mathrm{int}} \geq \sqrt{2}d$.

  As mentioned, this constraint's extent can be mitigated depending on the specific gate and its implementation.
  For instance, in the case of the CCZ gate \cite{levine_2019_parallel} in \Cref{fig:capabilities} (d), solely a single qubit needs to be within $r_{\mathrm{int}}$ of the others, while among themselves, they may be situated at a greater distance.
  On the other hand, depending on the implementation, constraints in addition to \Cref{eq:multi-qubit-mapping} must be fulfilled. 
  Taking again the example of the CCZ gate, one also has to take into consideration the geometric arrangement of the qubits. 
  In this case, the gate requires the three qubits to sit along a straight line, while a perpendicular arrangement would not be possible~\cite{levine_2019_parallel}.
\end{mapping}

\begin{scheduling}
  The restriction mechanism from the two-qubit gates generalizes to the case of two \mbox{multi-qubit} gates by composing the minimal distance $r_{\mathrm{re}}$ between any two qubits of the respective gates.
  So, for two gates $g(Q)$ and $g'(Q')$ acting on the two-qubit sets $Q, Q'$ we need
  \begin{equation}
    \label{eq:multi-qubit-scheduling}
    d(Q_{i},Q_{m}) \geq r_{\mathrm{re}} \quad \forall Q_{i} \in Q \text{ and } \forall Q_{m} \in Q' .
  \end{equation}
  In difference to the interaction constraint of \Cref{eq:multi-qubit-mapping}, this cannot be relaxed, as both control and target qubits are potentially in the Rydberg state during gate execution.
  As a result, multi-controlled gates restrict a larger number of atoms compared to simpler gates.
\end{scheduling}
The necessity for a larger interaction radius, along with the increased number of qubits, leads to an expanded restriction volume for multi-qubit gates in contrast to two-qubit gates.
Nevertheless, this could potentially be compensated by a more efficient synthesis process facilitated by utilizing a larger native gate set or the faster execution of the corresponding multi-qubit gate.

\subsubsection{Atom Shuttling}\label{sec:atom-shuttling}

\begin{figure*}[ht]
\centering
\includegraphics[width=0.95\textwidth]{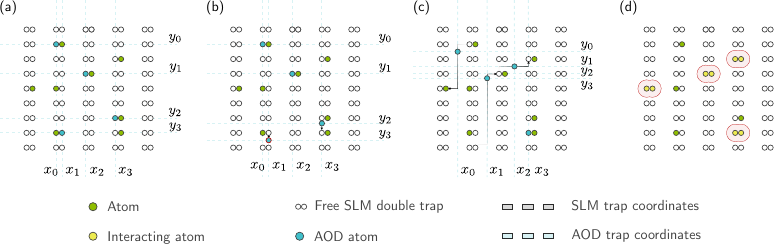}
\caption{
  \textbf{Neutral Atom Dynamically Field-Programmable Array (DPQA) Processor.} In this architecture, atoms are strategically placed within Spatial Light Modulator (SLM) trap pairs, ensuring that the distance between them is smaller than $r_{\mathrm{int}}$ while maintaining an inter-pair distance larger than $r_{\mathrm{re}}$ to prevent mutual restrictions.
  Computation in a DFPA processor can be considered as the repetition of three major phases: \textbf{(a) Loading phase:} A subset of atoms undergoes a switching process to be rearranged using Acousto-Optic Deflectors (AODs), indicated by the AOD coordinates $x$ and $y$.
  \mbox{\textbf{ (b + c) Shuttling phase:}} The trapped atoms are rearranged by readjusting the AOD coordinates.
  To avoid conflicts due to row crossing, the atom in the lower left corner (indicated in red) must be displaced first, allowing the other atom to occupy its designated position.
  In particular, for $y_2$ to reach its destination, the $y_3$ row has to be moved along as the coordinates are not allowed to cross.
  Subsequently, the remaining atoms, and also $y_3$, can be shuttled to their respective locations. 
  For more complicated shuttling operations this requires sophisticated methods to find adequate AOD movements.
  \textbf{(d) Gate phase:} An entangling CZ gate can be performed on atoms within the same pair of traps.
  This operation can be accomplished either with a global laser beam or individually using selective beams, as discussed in \Cref{fig:capabilities}.
  Phases (a) to (d) are repeated iteratively until all the required gates have been executed.
}
\label{fig:dpqa}
\end{figure*}

In the context of NAQC, individual tweezers holding qubits can be dynamically moved during computation without disrupting entanglement, as demonstrated in \cite{beugnon_2007_twodimensional,bluvsteinLogicalQuantumProcessor2023, bluvsteinQuantumProcessorBased2022a}.
This capability offers an alternative to implementing SWAP operations at the virtual level, requiring three CX gates.
In addition, physically shuttling atoms to new locations provides the flexibility to establish dynamic connectivity between qubits.

One physical realization of these shuttle operations involves placing atoms intended for shuttling within a tweezer array generated using a 2D crossed \emph{Acousto-Optic Deflector} (AOD).
On the contrary, stationary qubits remain in a static tweezer array formed by employing a \emph{Spatial Light Modulator} (SLM).
A typical complete shuttling operation requires first a pick-up from a static SLM to a dynamic AOD trap.
Via a controlled frequency ramp applied to the AOD, the qubit is rearranged to the destination, followed by a controlled release of the atom back into a static trap.
The selection of atoms to move can be altered for subsequent maneuvers, and specific parallel moves are feasible \cite{henriet_2020_quantum, kaufman_2021_quantum}, with the constraints summarized in the following box.

\begin{mapping}

  The AOD can be characterized by two sets of coordinates: ${x_{i}, \dots x_{k}}$ and ${y_{a}, \dots y_{c}}$.
  Each intersection $(x_{i},y_{a})$ defines a potential trap where an atom can be confined.
  Shuttling is achieved by modifying these coordinates, effectively relocating the corresponding traps.
  Consequently, changing the value of a specific coordinate, say $x_{0}$, results in the simultaneous movement of all AOD-trapped atoms in the first column, with identical displacements in the same direction.
  This implies that not all atoms within a single AOD can be moved independently.
  By examining the movement vectors $\vec{m}_{i,a}$ for atoms located at coordinate $(x_{i},y_{a})$, the following constraints emerge:

\begin{equation}
\begin{aligned}
\label{eq:shuttling_mapping}
\vec{m}_{i,a} \cdot \hat{e}_{x} &= \vec{m}_{i,b} \cdot \hat{e}_{x} \quad \forall a,b \\
\vec{m}_{i,a} \cdot \hat{e}_{y} &= \vec{m}_{j,a} \cdot \hat{e}_{y} \quad \forall i,j \; ,
\end{aligned}
\end{equation}

\noindent where $\hat{e}_{x}$ and $\hat{e}_{y}$ are the respective unit vectors.
An illustrative example of this scenario is shown in \Cref{fig:capabilities} (f), where two atoms are simultaneously shuttled row-wise. On the contrary, a third AOD-trapped atom remains stationary, corresponding to different row and column traps.

As a second constraint, we observe that two coordinates are not allowed to be closer than a given minimal distance $d_\mathrm{min}$.

\begin{equation}
\begin{aligned}
\label{eq:shuttling_crossing}
|x_{i} - x_{j}| &> d_\mathrm{min}  \quad \forall i,j \\
|y_{a}-y_{b}| &> d_\mathrm{min} \quad \forall a,b.
\end{aligned}
\end{equation}
Otherwise, this would result in overlapping trap potentials and undefined behavior with potential atom loss. \Cref{eq:shuttling_crossing} enforces that the coordinate lines cannot cross each other, maintaining a fixed ordering $x_{0}, \dots , x_{k}$ throughout the process.
\end{mapping}
\vspace{-10pt}
\begin{scheduling}
  There are two possible approaches to shuttle multiple AOD-trapped atoms in parallel.
  
  First, by spanning an AOD with multiple $x$ or $y$ coordinates to trap multiple atoms using a single AOD.
  In this scenario, all parallel movements must comply with the constraints described in \Cref{eq:shuttling_mapping} and \Cref{eq:shuttling_crossing}.
  Specifically, columns (rows) can only move along the same $x$ ($y$) displacement and must not cross each other.

  The second case involves using multiple AODs, each with their respective coordinates.
  In this setup, movements from different AODs are independent of each other, allowing for maximum flexibility.
  The remaining constraint is to ensure that the atoms keep a minimal distance through the shuttling process.
\end{scheduling}
\vspace{2cm}

Instead of using shuttling only as a substitute for virtual SWAP and MOVE operations, the recently discussed \emph{Dynamically Field-programmable Qubit Arrays} (DPQA or D-FPQA)~\cite{tanQubitMappingReconfigurable2022,bluvsteinLogicalQuantumProcessor2023,bluvsteinQuantumProcessorBased2022a,wangFPQACCompilationFramework2023} has emerged as a quantum processor design, which entirely focuses on shuttling.
Compared to the previous discussion, this new computation architecture is not based on qubits placed on a regular grid.
A short definition is given in the following box and described in \Cref{fig:dpqa}.
\begin{architecture}[DPQA]
  In the Dynamically Field-Programmable Qubit Arrays (DPQA) setup, stationary SLM traps are arranged in groups on a grid, ensuring that the qubit distance within a single group is smaller than $r_{\mathrm{int}}$, allowing gates to be applied within the group.
  Meanwhile, the groups themselves are placed at an inter-group distance greater than $r_{\mathrm{re}}$, preventing interference between gates on different groups, as illustrated in \Cref{fig:dpqa}.
  This arrangement requires more space than the previous configuration, especially for larger $r_{\mathrm{re}}$ values.

The computation process in DPQA can be divided into three phases, assuming that all atoms start in one of the SLM traps:
\begin{enumerate}
        \item \emph{Loading:} Depending on the following gates to be executed, atoms that need to be shuttled are loaded into the AOD traps by activating the corresponding coordinates.
  \item \emph{Shuttling:} Modulating the AOD coordinates enables the atoms to move column / row-wise.
        Due to the constraints of \Cref{eq:shuttling_mapping} and \Cref{eq:shuttling_crossing}, direct movement of atoms to their destinations may not always be possible, necessitating intermediate shuttling, as depicted in \Cref{fig:dpqa} (b).
        Here, the left atom has to be shuttled downwards first to let the other atom reach its destination.
        After reaching their destinations, the atoms switch traps back to stationary SLM traps.
  \item \emph{Gate execution:} Entangling gates can be performed on qubits within the same group.
        This can be achieved by addressing individual atoms, possibly implementing multi-qubit gates, or using a global beam to apply the same entangling gate on all groups with more than one atom.
        In this phase, required single-qubit gates can also be performed by addressing the atoms individually or globally.
\end{enumerate}
These three steps are iterated until all the gates in the circuit have been executed.
Due to the movement dependencies of \Cref{eq:shuttling_mapping}, finding suitable AOD movements is a highly non-trivial task.
\end{architecture}
An extension of the DPQA setup adds the possibility of defining separate zones for dedicated entangling, measuring and storing qubits which allows for experimentally optimized setups~\cite{bluvsteinLogicalQuantumProcessor2023}.
The routing between these zones on the other hand imposes an additional computational overhead, resulting in a trade-off situation regarding processor design similar to the discussions between gate-based and shuttling-based mapping in \Cref{sec:shuttling-numerics}.

\subsubsection{Measurements}\label{sec:measurements}

The qubit state of an atom can be measured by performing fluorescence imaging on a cycling transition between one of the computational states and an auxiliary electronic state \cite{saffman_2016_quantum}.
During this process, the fluorescence light emitted by the atoms is imaged with a camera such that bright spots in a final image correspond to atoms in a specific qubit state.
However, this procedure is experimentally challenging.
Scattering of photons for readout heats the atoms, leading to significant atom loss from the shallow optical dipole traps without laser cooling.
Measurements that preserve the atoms in the traps have been demonstrated in free space \cite{kwon_2017_parallel, covey_2029_2000times}, but might be easier to achieve by performing cavity-enhanced fluorescence imaging \cite{bochmann_2010_lossless, deistMidCircuitCavityMeasurement2022}.
Next to being faster and non-destructive, the latter technique has the advantage that much fewer photons must be scattered for detection, reducing atom heating and cross-talk problems at the expense of serial readout.
Parallel mid-circuit measurements of qubits have also been demonstrated recently \cite{graham_2023_midcircuit, norcia_2023_midcircuit, lis_2023_midcircuit, huie_2023_repetitive, singh_2023_midcircuit,bluvsteinLogicalQuantumProcessor2023}.

\subsubsection{Errors}\label{sec:errors}

Qubits in a quantum computation can suffer from \emph{idle-errors}, incurring when the qubits are unused, or \emph{gate errors} when imperfect quantum gates are applied.
During idling, NA qubits mainly experience two error processes.
First, atoms can decay from the qubit state with higher energy, commonly referred to as $\ket{1}$, to the ground state, which typically encodes the qubit $\ket{0}$ state.
This \emph{amplitude damping} causes the matrix element $\rho_{11}$ of the density matrix $\rho$ of a qubit to decay on a characteristic time scale $T_1$, called \emph{relaxation time}.
The relaxation time for NAQC is typically very large; it can be of the order of $\SI{100}{s}$ for hyperfine qubits, but even for other qubit encodings $T_1$ can reach several seconds~\cite{young_2020_halfminute}.
Second, atoms undergo \emph{dephasing}, mainly resulting from fluctuating external fields or drive fields.
Due to both error processes, the off-diagonal elements of the qubit's density matrix decay on a time scale $T_2^*$.
Canceling noise on slower time scales with dynamical decoupling techniques results in a modified, increased, \emph{dephasing time} $T_2$.
Encoding a qubit in two states insensitive to external field fluctuations yields long dephasing times that can be as high as several seconds.
As a simplification to the exact error analysis, which is highly non-trivial in general, we want to define the artificial parameter of an \emph{effective coherence time}
\begin{equation}
    \label{eq:t_eff}
    T_{\mathrm{eff}} = \frac{T_1 T_2}{T_1 + T_2} .
\end{equation}
If either $T_2 \ll T_1$ or vice versa, $T_{\mathrm{eff}}$ reduces to the respective shorter time, such that it correctly captures the typical time scale, limiting the system's coherence.

Furthermore, the application of gates can induce errors on a qubit.
Besides experimental imperfections such as e.g.~miscalibrations or laser intensity fluctuations, there are physical processes that limit achievable gate fidelities.
In particular, two- and multi-qubit gates require atoms to be temporarily excited to Rydberg states.
Those states decay quickly, either to the qubit subspace, leading to \emph{computational errors}, or to other electronic states, resulting in \emph{leakage errors}.
Experimentally, specific pulse shapes can be utilized to minimize the effect of such errors \cite{pagano_2022_error, jandura_2022_timeoptimal}.
Typically, these Rydberg decay errors are combined with other gate execution errors, resulting in a gate-specific \emph{average gate fidelity} $\mathcal{F}_g$ for a given gate $g$.

Unfortunately, even given the average gate fidelities of all gates in a quantum circuit, it is not possible to compute the exact final output fidelity.
In general, it is even hard to establish useful thresholds, as the exact fidelity depends on complicated microscopic error characteristics and their interplay with each other.
To still account for different gate fidelities and coherence times, we consider a strongly simplified error discussion in the following which should be considered only a very rough estimate of the actual errors.
Nonetheless, we can consider it a first proxy criterion regarding optimization techniques during compilation.

We can summarize this discussion of physical errors into two principal error types.
Decoherence errors for idle qubits and gate fidelities for executing gates.
In this simplified scheme, we can make the following abstraction from physical error processes to an error measure, which requires only a small set of physically motivated parameters to get a basic but hardware-aware error estimation.

\begin{errors}
  Given a quantum computation $U$ with synthesis, mapping, and scheduling already performed, including inserted MOVE and SWAP operations, we obtain a sequence of $\tilde{N}$  operations $O$:
  \begin{equation}
    \label{eq:sequence_of_operations}
    U = O_{N-1} \circ \dots \circ O_{\tilde{0}} . \nonumber
  \end{equation}
  In addition, let the fidelities $f$ of all operations and the coherence times $T_{1}$ and $T_{2}$ be given.
  We can define the abstract measure of \emph{approximate success probability} $P$. It is defined as the product of all average gate fidelities, combined with the term for the idle error:
  \begin{equation}
    \label{eq:success_prob}
    P(U) = \exp\left(-\frac{t_\mathrm{idle}}{T_{\mathrm{eff}}}\right) \prod_{i=0}^{\tilde{N}} \mathcal{F}_{O_i} ,
  \end{equation}
  \noindent where $T_{\mathrm{eff}}$ is given according to \Cref{eq:t_eff} and the \emph{idle-time} $t_\mathrm{idle}$ describes the time in which no gate is applied to a qubit, summed over the whole register.
  In the case that all operations, including SWAPs and MOVEs, are realized by gates, \Cref{eq:success_prob} simplifies to
  \begin{equation}
    \label{eq:success_prob_gate_only}
    P(U) = \exp\left(-\frac{t_\mathrm{idle}}{T_{\mathrm{eff}}}\right) \prod_{i=0}^{N} \mathcal{F}_{g_i} ,
  \end{equation}
  \noindent where we only need the gate fidelities $\mathcal{F}_(g_i)$ for all gates $g_i \in \Sigma_{\mathrm{native}}$ in the native gates set.\\
  The \emph{idle-time} can be computed given the gate execution times $t$ and the total circuit execution time $T$ as
  \begin{equation}
    \label{eq:idle_time}
    t_\mathrm{idle} = n \cdot T - \sum_{i=0}^{N} t(g_{i}) ,
  \end{equation}
  \noindent after routing has been employed to compute $T$.
\end{errors}

The scheme of approximate success probability provides a simple and fast-calculated proxy criterion for the occurrence of errors in quantum computation.
The general concept allows comparison between operations at different abstraction levels, such as native gates or oracles, and operations performed using additional capabilities, such as SWAP gates or SWAP shuttling movements, as long as it is possible to assign an operation-specific execution time $t$ and an average fidelity $\mathcal{F}$.

\subsubsection{Fault-tolerant QC and Error Correction}\label{sec:error-correction}
For the long-term goal of large-scale quantum computation, it will be necessary to employ techniques from quantum fault tolerance to deal with the accumulation of noise and errors~\cite{Campbell2017}.
In the case of NAs, not only computational errors but also leakage errors must be corrected to achieve fault tolerance.
There are proposals to convert leakage errors into Pauli $Z$-errors \cite{cong_2022_hardware} or detect and thus convert them into erasures \cite{wu_2022_erasure,sahay_2023_high,scholl_2023_erasure,ma_2023_highfidelity}.
Two-qubit gate designs have been proposed in which errors mainly occur in the form of detectable erasures \cite{jandura_2022_optimizing, fromonteil_2022_protocols}.

Regarding quantum error correction, most schemes require mid-circuit measurements and real-time feedback, a procedure which has been demonstrated lately with NAs \cite{singh_2023_midcircuit, huie_2023_repetitive,bluvsteinLogicalQuantumProcessor2023} but remains challenging.
Therefore, progress on measurement-free fault-tolerant quantum error correction might be promising for NAs \cite{heussen_2023_measurementfree,crowImprovedErrorThresholds2016,perlinFaulttolerantMeasurementfreeQuantum2023}.
Also, the possibility of using atom-photon interactions has been discussed lately~\cite{nagibRobustAtomphotonGate2023,liHighrateHighfidelityModular2024}, allowing for photonic interconnects on long distances.
Due to experimental demonstrations~\cite{bluvsteinLogicalQuantumProcessor2023}, current work is focused on the possibility of using the shuttling capability of NAs to prepare and manage encoded states for qLDPC~\cite{xuConstantOverheadFaultTolerantQuantum2023}, the surface code~\cite{viszlaiArchitectureImprovedSurface2023}, and generalized bicycle codes~\cite{viszlaiMatchingGeneralizedBicycleCodes2023}.
Combined with the discussion on fault-tolerant realizations of non-Clifford gates~\cite{wangEfficientFaulttolerantImplementations2023}, this paves the way towards fault-tolerant computation on the NAQC platform.
This is of particular interest, as the requirements for proposed qLDPC code realizations on superconducting platforms, are still beyond current hardware capabilities~\cite{delfosseBoundsStabilizerMeasurement2021, tremblayConstantOverheadQuantumError2022, strikisQuantumLowDensityParityCheck2023a}.

\section{Compilation for Neutral Atoms}\label{sec:compilation-for-na}
Meeting DiVincenzo's criteria at the hardware level is not sufficient to achieve a fully operational quantum computer.
Abstractly formulated quantum algorithms must be translated into physical operations available on the hardware and carefully coordinated to satisfy constraints and conditions imposed by the physical process or experimental setup.

In this section, we will explore how the computational capabilities and constraints of the NAQC platform, as discussed in \Cref{sec:comp-capabilities}, impact hardware-focused compilation steps.
Additionally, we will propose a set of simplified proxy criteria as figures of merit to evaluate the compilation results, enabling comparison between fundamentally different operations, such as virtual or physical swapping.

This discussion proposes potential compilation paths to address the increased number of possibilities, given the diverse quantum operation capabilities of NAs.
Such efforts may aid the development of design automation and compilation software, taking full advantage of the capability spectrum of the NAQC platform.
In particular, regarding the choice of figures of merit to evaluate the compilation results.

\subsection{Overview}\label{sec:overview-compilation}

\begin{figure*}[ht]
  \centering
  \includegraphics[width=\textwidth]{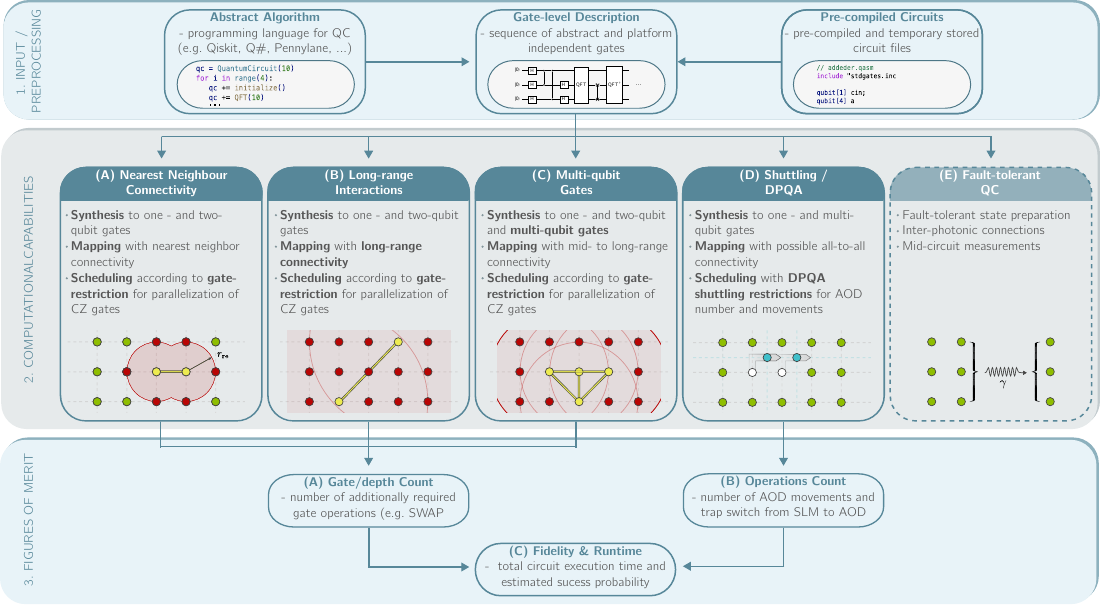}
  \caption{\label{fig:overview-compilation} \textbf{Compilation and Evaluation Process Overview for the Neutral Atom Quantum Computing Platform.} \mbox{\textbf{1. Input/Preprocessing:}}~The platform-independent compilation processes lead to a circuit description that includes abstract and hardware-independent gates.
    \textbf{2. Computational capabilities:} The different computational capabilities of the NAQC platform, as elaborated in \Cref{sec:comp-capabilities} with a short description of the constraints for the corresponding compilation subtasks of synthesis, mapping, and routing.
    Depending on the hardware setup, multiple capabilities, including all, can be considered.
    \textbf{3. Figures of merit:} The compilation output is evaluated based on capability-specific proxy criteria, such as gate or shuttling operation counts.
    To achieve a more comprehensive comparison and evaluation of gate-based routing and shuttling, figures of merit, such as the final execution time and fidelity of the compilation result, can be computed.
  }
\end{figure*}

The NAQC platform introduces a novel set of capabilities for implementing gates and new degrees of freedom for conducting operations, but consequently, also additional constraints within the compilation processes.
Moreover, for certain operations like shuttling, conventional figures of merit for evaluating compilation quality are no longer directly applicable.
This necessitates new schemes on how a given quantum algorithm is compiled and how the efficacy of the resulting compilation can be assessed.
To this end, we propose an abstract compilation overview that considers the distinct computational capabilities and appropriate figures of merit for optimization processes and evaluation in the specific case of NAs.
The proposed procedure encompasses three main steps, depicted in \Cref{fig:overview-compilation} as a three-layer diagram.

\begin{enumerate}[wide, labelindent=0pt]
  \item \emph{Input/pre-processing:} The initial step involves providing the input to the workflow, comprising a quantum circuit in which all \mbox{platform-independent} optimizations have already been performed.
  \item \emph{Computational capabilities:} This part involves the compilation process, which varies depending on the specific hardware capabilities, encompassing long-range interactions, multi-qubit gates, and shuttling as discussed in \Cref{sec:comp-capabilities}.
  \item \emph{Figures of merit:} The final step requires using suitable figures of merit to evaluate the quality of the compilation, which may differ across various capabilities.
\end{enumerate}

\subsubsection{Input/Pre-processing}\label{sec:input-compilation}
The initial stage acts as the input layer in the proposed compilation overview. 
It outputs a fully optimized circuit where all hardware- and platform-independent optimizations have been applied.
Because this phase is not tied to any particular platform, well-established optimization techniques and software initially designed for SC hardware can be employed here. 
The result is a sequence of gates, still including abstract gates that will be broken down in the subsequent steps based on the available hardware. 
Certain aspects of this process can also be pre-compiled beforehand, stored, and later retrieved from an intermediate representation \cite{OpenQASMBroaderDeeper2023,lubinskiAdvancingHybridQuantum2022}.

\subsubsection{Computational Capabilities}\label{sec:capab--constr}
The subsequent layer includes hardware-dependent compilation steps, specifically \emph{synthesis}, \emph{mapping}, and \emph{scheduling}.
Unlike the previous optimizations, these steps depend on the available operations, such as the basis gate set and how SWAP operations can be performed.
Consequently, the compilation process may vary based on the set of available capabilities and the corresponding constraints, which directly correspond to the sections of \Cref{sec:comp-capabilities}.

\begin{enumerate}[wide, labelindent=0pt,label=(\Alph*)]
\item \emph{Nearest-neighbor connectivity:} \
        The first option involves utilizing capabilities equivalent to those often considered for the SC platform.
        This includes employing single-qubit rotations and Rydberg blockade-based controlled phase gates, which differ from CX by just single-qubit operations.
        Together, they form a universal gate set for decomposing non-native gates during synthesis.
        Qubit connectivity follows nearest-neighbor connections, defining the connectivity graph $G$.
        Synthesis, mapping, and routing techniques applied in SC systems can be adapted with adjustable parameters such as gate time or fidelity.
        The additional constraint, \emph{gate restriction} from \Cref{sec:2qubitgates}, must be considered on top as it is unique to the NAQC platform.

\item \emph{Long-range interactions:} \
        One significant distinction of NAs is their ability to execute two-qubit phase gates not only between adjacent qubits but also between any two atoms where the blockade interaction reaches sufficient strength.
        This increases connectivity, reducing the need for additional SWAP operations during circuit execution.
        However, the larger restriction volume due to long-range interactions may lead to more sequential gate execution.
\item \emph{Multi-qubit gates:} \
        Besides higher connectivity, the long-range Rydberg interaction enables the implementation of native multi-qubit gates, which is particularly beneficial for the synthesis task or algorithms that inherently consist of many multi-qubit gates, such as reversible classical logic circuits.
        However, involving more atoms also increases the restriction volume, potentially limiting simultaneous gate executions.

\item \emph{Shuttling/DPQA:} \
        A fundamental advantage of the NAQC platform is the ability to physically move atoms, and hence qubits, rather than relying on virtual SWAP operations.
        High-fidelity shuttling is a promising alternative for gates between distant qubits.
        Additionally, a fully shuttling-based architecture (dynamically field-programmable quantum array - DPQA) might be possible, eliminating the need for virtual swaps altogether.

\item \emph{Fault-tolerant quantum computing:} \
        NAs also offer additional capabilities relevant to future compilation tasks, such as fault-tolerant quantum computing.
        The details of compiling logical circuits are beyond the scope of this work, but with the recent experimental progress, first work in this direction has been done, in particular discussions on the implementation of qLDPC codes~\citet{xuConstantOverheadFaultTolerantQuantum2023}, surface codes~\cite{viszlaiArchitectureImprovedSurface2023}, and generalized bicycle codes~\cite{viszlaiMatchingGeneralizedBicycleCodes2023}.

\end{enumerate}
Theoretically, it is feasible to combine multiple, or even all, of the aforementioned capabilities. For instance, it is possible to harness both physical shuttling operations and virtual SWAP gates within the same computation.
Alternatively, one could adopt the DPQA approach while organizing atoms into subgroups instead of pairs, thereby utilizing shuttling and the ability to implement multi-qubit gates within the atomic subgroups.
Examples are given by \citet{viszlaiArchitectureImprovedSurface2023} studying possible surface code architectures to take advantage of such a setup, or mapping algorithms taking advantage of using both, gate and shuttling-based mapping techniques~\cite{schmidHybridCircuitMapping2023}.
Nevertheless, there remains a diverse set of compilation tasks and open avenues for further investigation.

\subsubsection{Figures of Merit}
\label{sec:performance-measure}
The evaluation of a compilation output requires the use of appropriate performance measures or metrics.
The choice of metrics varies depending on whether the aim is to compare different compilers, capabilities, or platforms.
For instance, the common practice of counting additional swap gates as a measure in SC compilers lacks significance for a DPQA-based architecture that avoids swap gates altogether.
Therefore, novel metrics may be required to measure the quality of a compilation output for NAs.
The same holds for the figures of merit used during optimization in the form of cost functions.
These figures of merit, suitable for NA-based hardware, are discussed in the following.

\begin{enumerate}[wide, labelindent=0pt,label=(\Alph*)]
        \item \emph{Gate count:}
        The predominant approach to evaluate the quality of a compilation is to quantify the total number of gates.
        Directing attention to the gates with the lowest fidelity often suffices, as they exert the most significant influence on the final output.
        In the context of synthesizing, notable metrics include the counts of CX gates for NISQ computing or T gates in the regime of fault-tolerant quantum computing.
        In the case of NAs, a similar figure of merit could be employed, where, depending on the hardware, the focus lies on the gates with the lowest fidelity or longest execution time.

        \noindent In addition to the total number of gates, another critical aspect to consider is the number of gate layers, referred to as \emph{circuit depth}, where each layer contains operations that can be executed simultaneously. This evaluation considers the scheduling of operations and serves as a proxy criterion for estimating the overall execution time.
        However, when dealing with NA, especially in cases involving nearest-neighbor interactions (2.A), it is imperative to account for the existence of gate restriction volumes, as elaborated in \Cref{sec:2qubitgates}, during the scheduling process. This will result in larger circuit depths in general.

        \item \emph{Operation count:}
        Since shuttling does not introduce additional gates into the circuits, the previous gate-based measure is not applicable.
        An alternative approach would be to define a similar and straightforward operation count.
        A suitable candidate for this count could be the number of AOD-based MOVE or SWAP operations.
        For DPQA architectures, one can use the number of iterations of the three phases of loading, shuttling, and execution as a first-order evaluation measure.

        \item \emph{Fidelity and runtime:}
        The diverse possibilities for implementing a quantum circuit on the NAQC platform render simple measures such as gate count less suitable, especially when multiple capabilities are leveraged.
        The most comprehensive measure for evaluating a quantum computation is the output fidelity and the total runtime.
        Unfortunately, computing the exact output fidelity is generally not feasible.
        To address this limitation, we considered the proxy of \emph{approximate success probability} of \Cref{eq:success_prob}, quantifying the likelihood of executing the circuit successfully without errors.
        It enables a comparative analysis of compilation outcomes for various circuits, facilitating the evaluation of a given compiler's performance.
        Additionally, this metric allows for comparing different capabilities, assisting in estimating the most promising approach for mapping, for example, whether employing nearest neighbor, long-range SWAP gates, or shuttling SWAP operations are most suitable.

\end{enumerate}

\subsection{Compilation Parameters}
\label{sec:comp-param}
All stages within the overview depicted in \Cref{fig:overview-compilation} are contingent upon hardware-specific parameters, such as the strength of the Rydberg blockade, the gate execution time for the mapping and -scheduling task, and the corresponding gate fidelities used to compute the final output fidelity.
These parameters heavily depend on the chosen hardware and the overall experimental configuration, including the atom species and the protocol employed to implement specific gates.
Nonetheless, the development of compilers necessitates the utilization of reasonable hardware parameters to generate practical results for hardware experts and other users.
Hence, we present a list of estimated parameters in \Cref{tab:parameters} for two exemplary setups.
These parameters align with the more abstract description of the physical capabilities, suitable for tool developers, and discussed in \Cref{sec:comp-capabilities}.
It is important to note that these parameters are not fixed and will inevitably evolve in the coming years with advancements in hardware and control systems.
The underlying idea is that one can then reason on an updated set of the same or similar parameters but with the same general considerations, as all shown parameters can change with future hardware improvements.
Therefore, the objective is to develop more adaptable compilers that can adjust to specific hardware configurations provided.

\begin{table*}[ht]
  \centering
\begin{threeparttable}
\begin{tabular}{@{}lllll@{}}
\toprule
                         &                     &                  & \textbf{Setup 1}     & \textbf{Setup 2}     \\ \midrule
\textbf{General}         & Atom species        &                  & Strontium ~\cite{young_2020_halfminute,madjarov_2020_highfidelity,shawMultiensembleMetrologyProgramming2024}           & Rubidium~\cite{bluvsteinQuantumProcessorBased2022a,evered_2023_highfidelity}             \\
                         & Inter atomic distance & $d$              & $\SI{3}{\micro\m}$  &  $\SI{3}{\micro\m}$   \\ \midrule
\textbf{Gate durations} $t$  & Single-qubit     &       & $\SI{200}{\micro\s}$~\cite{shawMultiensembleMetrologyProgramming2024}                   & $\SI{0.5}{\micro\s}$~\cite{Levine2022}                  \\
                         & CZ                  &                  & $\SI{0.1}{\micro\s}$~\cite{madjarov_2020_highfidelity} & $\SI{0.2}{\micro\s}$~\cite{evered_2023_highfidelity} \\
                         & CCZ                 &                  & $\sim \SI{1}{\micro\s}$ & $\sim \SI{1}{\micro\s}$   \\
                         & CCCZ                 &                  & $\geq\SI{1}{\micro\s}$   & $\geq\SI{1}{\micro\s}$   \\ \midrule
\textbf{Gate fidelities} $\mathcal{F}$ & Single-qubit         &                  & $> 0.99$~\cite{shawMultiensembleMetrologyProgramming2024}             & $> 0.999$~\cite{Levine2022}            \\
                         & CZ                  &                  & $> 0.99$             & $> 0.995$~\cite{evered_2023_highfidelity}            \\
                         & CCZ                 &                  & $\sim 0.95$               & $> 0.98$~\cite{evered_2023_highfidelity}             \\
                         & CCCZ                 &                  & $\sim 0.95$               & $\sim 0.95$               \\ \midrule
\textbf{Coherence times} & Qubit decay     & $T_1$ & $> \SI{1}{\s}$~\cite{young_2020_halfminute}                         & $> \SI{100}{\s}$                       \\
                         & Dephasing           & $T_2^{\,\S}$            & $> \SI{10}{\s}$~\cite{young_2020_halfminute}      & $> \SI{1.5}{\s}$~\cite{evered_2023_highfidelity}     \\ \midrule
\textbf{Long-range}      & Interaction radius  & $r_\mathrm{int}$ & 2                    & 2                    \\
                         & Restriction radius  & $r_\mathrm{re}$  & 4                    & 4                    \\ \midrule
\textbf{Shuttling}       & Fidelity            &                  & 1          & $1^\ast$~\cite{bluvsteinQuantumProcessorBased2022a}          \\
                         & Shuttling speed & $v_s$ & $< \SI{0.025}{\micro\m\per \micro \s} $~\cite{shawMultiensembleMetrologyProgramming2024} & $< \SI{0.55}{\micro\m\per \micro \s} {}^\ast$~\cite{bluvsteinQuantumProcessorBased2022a} \\
                         & Trap swapping       &                  & $\SI{20}{\micro\s}$~\cite{beugnon_2007_twodimensional}  & $\SI{20}{\micro\s}$~\cite{beugnon_2007_twodimensional} \\
\bottomrule
\end{tabular}
\begin{tablenotes}\footnotesize
\item [$\ast$] According to Fig. 1d of \citet{bluvsteinQuantumProcessorBased2022a} we assume perfect fidelity if the average shuttling speed is below the indicated threshold.
\item [$\S$] We consider $T_2$ times using additional dynamic decoupling techniques.
\end{tablenotes}
\end{threeparttable}

\caption{The hardware parameters pertain to the compilation steps of mapping and scheduling, alongside the subsequent evaluation measures shown in the overview of \Cref{fig:overview-compilation}.
  These parameters are provided for two hypothetical experimental setups and are intended solely for a preliminary estimation, as they are dependent on the specific hardware utilized and are likely to change with future experimental improvements.}
\label{tab:parameters}
\end{table*}

\subsection{Discussion}
\label{sec:flow-discussion}
\Cref{fig:overview-compilation} provides a comprehensive overview of the whole hardware-dependent compilation process by bringing the summary boxes of \Cref{sec:comp-capabilities} in relation to each other and to possible figures of merit for evaluation.
Nevertheless, it should not be viewed as an inflexible framework dictating the compilation process.
Instead, it should be perceived as a graphical representation delineating the broad spectrum of compilation possibilities of the NAQC platform.
This complexity arises from the potential selection or amalgamation of multiple capabilities to achieve improved outcomes.
In addition to the increased number of compilation options, the overview underscores the need for additional or more intricate metrics and figures of merit, depending on the available capabilities.
Depending on the available hardware, a compiler can take multiple paths along the three layers, leveraging different capabilities.
As a tool developer, one must check which paths the compiler should cover and, therefore, which figures of merit are suitable for optimization.
In the following \Cref{sec:rel-work-software}, we will consider already available examples of compilers and discuss which path in \Cref{fig:overview-compilation} has been chosen.
Afterward, in \Cref{sec:numer-eval}, we will perform multiple selected case studies to compare error rates along different paths and discuss appropriate figures of merit, depending on the given hardware parameters.

\section{Related work and software}\label{sec:rel-work-software}
Based on the compilation overview of \Cref{fig:overview-compilation}, we want to give a concise summary of already available compilation software for NAs.
For this aim, we categorized the compilers based on the capability they focus on most, including long-range, multi-qubit gates, and shuttling.
Additionally, we also discuss the usage of SC compilers for the NAQC platform, since we will employ such an approach for some of the case studies in \Cref{sec:numer-eval}.
We briefly discuss each compiler's functionality and the corresponding methods used.
All compilers and the respective supported capabilities are summarized in \Cref{tab:compiler_overview}.
This section aims to review currently available compilers and aids interested tool developers in identifying subproblems that still lack a corresponding software solution.

\begin{table*}[htp]
\begin{threeparttable}

  \begin{tabular}{@{}lccccccccccc@{}}
\toprule
\textbf{Capabilities} & \rot{SC~\cite{Qiskit,bergholmPennyLaneAutomaticDifferentiation2022,developersCirq2023}} & \rot{\citet{bakerExploitingLongDistanceInteractions2021a}} & \rot{Q-Tetris \cite{liTimingAwareQubitMapping2023}} & \rot{Geyser \cite{patelGeyserCompilationFramework2022a}} & \rot{\citet{brandhoferOptimalMappingNearTerm2021a}} & \rot{OLSQ-DPQA~\cite{tanCompilingQuantumCircuits2023}} & \rot{\citet{nottinghamDecomposingRoutingQuantum2023}} & \rot{\citet{schmidHybridCircuitMapping2023}} & \rot{FPQA-C~\cite{wangFPQACCompilationFramework2023}} & \rot{Q-Pilot~\cite{wangQPilotFieldProgrammable2023}} & \rot{\citet{tanDepthOptimalAddressing2D2024}} \\ \midrule
\textbf{Long-range} &  &  &  &  &  &  &  & &  & \\
Synthesis & \faCheckSquareO &  &  & \faCheckSquareO\rlap{$^{\ast}$}  &  & \faCheckSquareO\rlap{$^{\ast}$}  & \faCheckSquareO\rlap{$^{\ast}$} & \faCheckSquareO\rlap{$^{\ast}$} & \faCheckSquareO\rlap{$^{\ast}$} & \faCheckSquareO\rlap{$^{\ast}$} & \\
\begin{tabular}[c]{@{}l@{}} Mapping \& \\ Routing\end{tabular} & \faCheckSquare & \faCheckSquare & \faCheckSquare & \faCheckSquareO\rlap{$^\S$} & \faCheckSquareO\rlap{$^\dagger$} &  & \faCheckSquareO\rlap{$^{\ddagger}$} & \faCheckSquare & \faCheckSquare\rlap{$^{\S}$} &  &\\
Scheduling &  & \faCheckSquare & \faCheckSquare & \faCheckSquareO &  & \multicolumn{1}{l}{} & \faCheckSquare & \faCheckSquare & \faCheckSquare & & \\ \midrule
\textbf{Multi-qubit} &  &  &  &  &  &  &  & & & & \\
Synthesis &  &  &  & \faCheckSquare\rlap{$^{\ast\ast}$} &  &  &  & & & & \\
\begin{tabular}[c]{@{}l@{}} Mapping \& \\ Routing\end{tabular} & \faCheckSquareO & \faCheckSquare & \faCheckSquare & \faCheckSquareO &  &  &  & \faCheckSquare & & & \\
Scheduling &  & \faCheckSquare & \faCheckSquare & \faCheckSquareO &  &  &  & \faCheckSquare & & & \\ \midrule
\textbf{Shuttling} & \multicolumn{1}{l}{} & \multicolumn{1}{l}{} & \multicolumn{1}{l}{} & \multicolumn{1}{l}{} & \multicolumn{1}{l}{} & \multicolumn{1}{l}{} &  & & & & \\
Mapping &  &  &  &  & \faCheckSquare\rlap{$^{\S\S}$} & \faCheckSquare & \faCheckSquare & \faCheckSquare & \faCheckSquare & \faCheckSquare & \\
Scheduling &  &  &  &  & \faCheckSquare & \faCheckSquare & \faCheckSquare & \faCheckSquare & \faCheckSquare& \faCheckSquare & \\ \midrule
\textbf{Single-qubit} & \multicolumn{1}{l}{} & \multicolumn{1}{l}{} & \multicolumn{1}{l}{} & \multicolumn{1}{l}{} & \multicolumn{1}{l}{} & \multicolumn{1}{l}{} &  & & & & \\
Synthesis &  &  &  &  &   &  & \faCheckSquare & & & &  \faCheckSquare \\
Scheduling & \faCheckSquareO  &\faCheckSquareO  &\faCheckSquareO  &\faCheckSquareO  &\faCheckSquareO   &  & \faCheckSquare &\faCheckSquareO &\faCheckSquareO &\faCheckSquareO & \faCheckSquare \\ \midrule
\textbf{Setup} & indiv.  & indiv.  & indiv.  & indiv.  & indiv. & DPQA  & indiv.  & indiv. & DPQA & DPQA & - \\ \midrule
\textbf{General} & \multicolumn{1}{l}{} & \multicolumn{1}{l}{} & \multicolumn{1}{l}{} & \multicolumn{1}{l}{} & \multicolumn{1}{l}{} & \multicolumn{1}{l}{} &  & & & & \\
\begin{tabular}[c]{@{}l@{}}Technique/\\ Algorithms\end{tabular} & \begin{tabular}[c]{@{}c@{}}SMT, A*, \\ heuristics, \\ ....\end{tabular} & \begin{tabular}[c]{@{}c@{}}look-ahead \\ heuristic\end{tabular} & \begin{tabular}[c]{@{}c@{}}heuristic,\\ MCTS\end{tabular} & heuristic & SMT & \begin{tabular}[c]{@{}c@{}}SMT, \\ heuristic\end{tabular} & heuristic & heuristic & \begin{tabular}[c]{@{}c@{}}MAX k-Cut,\\ heuristic \end{tabular} & heuristic & SMT \\
\begin{tabular}[c]{@{}l@{}}Open source \\ availability\end{tabular} & \begin{tabular}[c]{@{}c@{}}Github, \\ PiPI, ...\end{tabular} & \begin{tabular}[c]{@{}c@{}}Github\\ \cite{littekenNeutralAtomCompilation2023}\end{tabular} & \begin{tabular}[c]{@{}c@{}}Github\\ \cite{QTetris2022}\end{tabular} & \begin{tabular}[c]{@{}c@{}}Zenodo\\ \cite{tirthak_patel_2022_7084132}\end{tabular} & - & \begin{tabular}[c]{@{}c@{}}Github\\ \cite{OLSQDPQACompiler2023}\end{tabular} & - & \begin{tabular}[c]{@{}c@{}}Github\\ \cite{LsschmidMqtqmapHybridmapper} \end{tabular} & - & - & - \\
Language & \begin{tabular}[c]{@{}c@{}}Python, \\ \CC, Rust, \\ ....\end{tabular} & Python & \CC & Python &  & Python &  & \CC & & & \\
\begin{tabular}[c]{@{}l@{}}General input \\ (e.g.~QASM)\end{tabular} & \faCheckSquare & \faCheckSquareO & \faCheckSquareO & \faCheckSquareO &  &\faCheckSquareO  & & \faCheckSquare & & & \\ \bottomrule
\end{tabular}

\begin{tablenotes}\footnotesize
\item [$\ast$] conversion to native gate set (CZ-gates).
\item [$\S$] uses mapping from~\cite{Qiskit}
\item [$\dagger$] uses mapping from~\cite{tanOptimalLayoutSynthesis2020}.
\item [$\ddagger$] uses mapping from~\cite{cowtanQubitRoutingProblem2019}.
\item [$\ast\ast$] Composition of Toffoli gates.
\item [$\S\S$] only 1D displacement of grid row/columns.
\end{tablenotes}
\end{threeparttable}
\caption{\textbf{Overview of Presently Available Compilers for the NA Platform.}
  The symbol \faCheckSquare \space indicates the main functionality of the software and implies full support, while \faCheckSquareO \space refers to partial support only or reliance on external software. The setup differentiates between compilers that assume individual addressability of entangling gates and the DPQA setup discussed in \Cref{sec:atom-shuttling}.}
\label{tab:compiler_overview}
\end{table*}

\subsection{Compilers for Superconducting Platform}
\label{sec:superc-comp}
Before discussing compilers specific to the NAQC platform, we first want to briefly discuss the idea of reusing already available compilers.
While general compilation for NAQC has to consider a more extensive range of capabilities and constraints, it is still possible to leverage e.g. currently available SC compilers with additional pre- and post-processing steps to get compilation outputs valid for NAQC.
In this way, the full potential of NA-specific capabilities cannot be fully exploited. However, we still want to mention this possibility here, as we used similar techniques for the evaluations in \Cref{sec:numer-eval}.
Furthermore, SC compilation algorithms may provide a suitable basis for generalization and adaptation to the NAQC platform.

The SC platform has access to a broad range of advanced software tools \cite{Qiskit,bergholmPennyLaneAutomaticDifferentiation2022,developersCirq2023,steigerProjectQOpenSource2018,willeMQTQMAPEfficient2023,tanOptimalLayoutSynthesis2020,sivarajahKetRetargetableCompiler2020} that greatly facilitate the compilation process.
However, when applied to the NAQC platform, these tools are only partially suitable for the synthesis, mapping, and scheduling steps.

\begin{itemize}
\item\emph{Synthesis}: As the SC platform lacks straightforward support for multi-qubit gates, the corresponding synthesis steps usually decompose to one- and two-qubit gates, omitting capability $\text{2.C}$ from \Cref{fig:overview-compilation}.

\item\emph{Mapping}: SC relies on gate-based virtual swapping to establish the required connectivity, excluding $\text{2.D}$ from the process. However, the mapping problem can be adapted for NAs by defining the coupling graph $\mathbf{C}$ based on the interaction radius $r_{\mathrm{int}}$ as shown in \Cref{eq:coupling_graph}.

\item\emph{Scheduling}: Additional scheduling constraints of the restriction volume according to \Cref{eq:blocking} require an extra scheduling post-processing step for the NAQC platform.
\end{itemize}

Evaluation proxies commonly used in these tools involve gate count and circuit depth.
This leads to the following path in \Cref{fig:overview-compilation}: $1 \rightarrow \text{2.A/2.B} \rightarrow \text{3.A}$.

\subsection{Long-range Compiler}
\label{sec:restr-aware-comp}

When directly compared to other QC platforms, one of the constraints of the NAQC platform is the occurrence of restriction volumes whenever an atom employs the Rydberg state to execute a two- or multi-qubit gate.
This significantly impacts the scheduling task, as such gates that restrict each other must be executed sequentially rather than in parallel, increasing the total execution time.
Given that quantum information can only be stored for a limited time without decoherence, this directly affects the output fidelity, which can be estimated according to \Cref{eq:success_prob}.

\textbf{1.} The initial solution to address this problem was proposed by \citet{bakerExploitingLongDistanceInteractions2021a}, with an openly available Python package \cite{littekenNeutralAtomCompilation2023}.
Their approach assumes that the synthesis task has already been performed and concentrates on the mapping part and potential strategies to mitigate atom loss.

\begin{itemize}
\item \emph{Mapping}: Utilizing a look-ahead scheme to select the shortest SWAP path with minimal disruption for future interactions.
\item \emph{Scheduling}: Execute swap gates in parallel when they do not impose restrictions on other operations.
\end{itemize}

\citet{bakerExploitingLongDistanceInteractions2021a} primarily investigated the trade-off between long-range interactions and the effect of the simultaneously increasing restriction volume.

\begin{remark}
  It should be noted that Baker \textit{et al.} consider a slightly different definition of the restriction volume than the one proposed in our work.
  Additionally, they assume the restriction to be variable, depending on the inter-qubit distance of the gate and, therefore, possibly varying from gate to gate.
\end{remark}

The Mapper also supports multi-qubit gates, enabling the utilization of the full gate-based mapping capabilities of the NAQC platform, represented as $1 \rightarrow \text{2.A-C} \rightarrow \text{3.A}$ in \Cref{fig:overview-compilation}.

\textbf{2.} Addressing the same problem formulation, \citet{liTimingAwareQubitMapping2023} presented a \CC \space based solution named \emph{Q-Tetris}, inspired by the resemblance of the problem to the renowned block puzzle game.
In contrast to Baker \textit{et al.}, they consider distinct execution times for single, multi-qubit, and SWAP gates.

\begin{itemize}
\item \emph{Mapping \& scheduling}: Employ a greedy heuristic algorithm in conjunction with a Monte Carlo tree search (MCTS) approach to strategically insert necessary SWAP gates and simultaneously minimize the overall circuit execution time.
\end{itemize}

By arranging the gates in this time-aware manner, they claim that their method produces more time-efficient circuits compared to Baker \textit{et al.} Hence, they do not consider the gate count but the total execution time of the circuit.
Nevertheless, their approach supports the same capabilities as Baker \textit{et al.}, leading to the compilation path $1 \rightarrow \text{2.A-C} \rightarrow 3.A \rightarrow 3.C$ in \Cref{fig:overview-compilation}.

\subsection{Multi-qubit Compiler}
\label{sec:multi-qubit-compiler}
Although both previous compilers only support the mapping and scheduling tasks, as of our current knowledge, there is no available software capable of performing multi-qubit-aware synthesis for general quantum algorithms.
For classical reversible logic, tools~\cite{zulehnerOnePassDesignReversible2018,soekenSynthesisReversibleCircuits2012} exist that create circuits containing Toffoli and higher controlled NOT gates.

\textbf{1.} However, ~\citet{patelGeyserCompilationFramework2022a} and the corresponding open-source Python software \emph{Geyser}\cite{tirthak_patel_2022_7084132} take a different approach.
Instead of decomposing a large unitary into a basis gate set containing multi-qubit gates, they compose blocks of Toffoli gates from a set of two-qubit gates.

\begin{itemize}
  \item\emph{Synthesis}: The approach involves identifying gate blocks containing three qubits and seeking an equivalent gate sequence built from Toffoli gates and single-qubit gates.
        This is achieved by minimizing the Hilbert-Schmidt distance between the original gates and the substitute gates.
  \item\emph{Mapping}: The SC compiler Qiskit~\cite{Qiskit} is used, considering that the constructed Toffoli blocks do not restrict each other.
\end{itemize}

For evaluation, they do not consider gate count but rather laser pulse count, which can be seen as an estimation for the total execution time of the circuit.
In terms of the overview in \Cref{fig:overview-compilation}, this leads to a similar path as the previous two compilers, $1 \rightarrow \text{2.A-C} \rightarrow \text{3.A} \rightarrow \text{3.C}$, but with a focus on the previously unexplored synthesis task, allowing the possibility of leveraging both approaches.

\subsection{Shuttling Compiler}
\label{sec:shuttling-compiler}
All previous compilers have solely considered the virtual gate-based SWAP operation for the mapping task.
However, the Rydberg platform offers the additional capability of physically rearranging atoms to achieve an equivalent SWAP or a simple MOVE operation.

\textbf{1.} The pioneering attempt to harness this extra degree of freedom was made by \citet{brandhoferOptimalMappingNearTerm2021a}.
Their approach does not encompass general shuttling operations but focuses solely on one-dimensional displacements of atom rows.
This choice is justified because atoms trapped in rows or columns using AOD traps can be efficiently shuttled in parallel.

\begin{itemize}
  \item \emph{Mapping}: The model involves considering SWAP gate insertion (similar to SC~\cite{tanOptimalLayoutSynthesis2020,willeMappingQuantumCircuits2019}) and potential one-dimensional displacements.
        The authors employ Satisfiability Modulo Theories (SMT)~\cite{biere_handbook_2009} solvers to find an optimal solution.
\end{itemize}

They compute the final circuit fidelity to evaluate the results and compare gate-based swapping with displacement-based swapping.
Consequently, the compiler path in \Cref{fig:overview-compilation} can be described as $1 \rightarrow \text{2.A/B} \rightarrow \text{3.A} \rightarrow \text{3.C}$ and $1 \rightarrow \text{2.D} \rightarrow \text{3.B} \rightarrow \text{3.C}$, covering a significant portion of the NA capabilities.

Rather than considering only one-dimensional displacements, ~\citet{tanCompilingQuantumCircuits2023} focus on a shuttling-only DPQA architecture as discussed in \Cref{sec:atom-shuttling}.
The corresponding open-source Python package~\cite{OLSQDPQACompiler2023} enables users to explore DPQA movements and offers helpful visual animations of these operations.

\begin{itemize}
  \item \emph{Mapping \& scheduling}: They encode the possible DPQA movements as an SMT problem and solve it to optimize the number of shuttling operation cycles.
        Additionally, they propose a heuristic approach based on the exact solver with better runtime scaling.
\end{itemize}
In the overview \Cref{fig:overview-compilation}, this compiler corresponds to the path $1 \rightarrow \text{2.D} \rightarrow \text{3.B}$, but this time it takes advantage of the full two-dimensional shuttling capability.

\textbf{2.} Recently, \citet{nottinghamDecomposingRoutingQuantum2023} presented a novel compiler that addresses the crucial synthesis step for the NAQC platform, considering the global addressing of the qubits.
Additionally, they propose a unique mapping approach for a shuttling-only architecture, with less stringent constraints compared to \citet{tanCompilingQuantumCircuits2023}.

\begin{itemize}
  \item \emph{Synthesis}: The authors propose two distinct decomposition approaches to address the challenge posed by single-qubit rotations that are only available as global beams.
        They leverage Qiskit's internal decomposition to single-qubit and CZ gates and apply their own optimization post-processing.
  \item \emph{Scheduling}: Accounting for global beams in single-qubit gates, the gates are organized into groups of parallelizable single-qubit gates and parallelizable CZ gates.
        CZ gates that restrict each other are executed sequentially.
  \item \emph{Mapping}: By relaxing the shuttling constraints from \Cref{eq:shuttling_mapping} and \Cref{eq:shuttling_crossing}, they allow different AOD rows/columns to cross.
        This enables the definition of a \emph{movement graph} that indicates all possible shuttling moves analogous to the connectivity graph for gate-based interactions.
        Mapping is then performed on this denser graph by selecting shuttling movements that bring the gate qubits closer to each other.
\end{itemize}

\citet{nottinghamDecomposingRoutingQuantum2023} marks the first step toward synthesis for NAs, addressing global single-qubit rotations and CZ gates.
They subsequently evaluate both virtual swapping and shuttling-based swapping in a DPQA-like problem scenario.
Their evaluation accounts for the influence of a gate-based error model and decoherence errors.
Regarding the compilation overview, this approach can be described as $1 \rightarrow 2.A/B/D \rightarrow 3.A/B \rightarrow 3.C$, encompassing all NA-specific capabilities except for native multi-qubit gate support.

\textbf{3.} Supporting all capabilities of \Cref{fig:overview-compilation} was done by \citet{schmidHybridCircuitMapping2023}, proposing a hybrid compilation scheme where a heuristic decides the most suitable mapping capability for each gate.
\begin{itemize}
  \item \emph{Mapping}: A SABRE-based~\cite{liTacklingQubitMapping2019} heuristic is used to compute the necessary SWAP gates. The duration and fidelity of these are compared to AOD-based shuttling operations, choosing the more favorable option.
  \item \emph{Scheduling}: Assumes single addressability and taking into account multi-qubit gate restrictions. Shuttling operations are converted to the respective AOD movements and scheduled according to the constraints in \Cref{eq:shuttling_mapping} and \Cref{eq:shuttling_crossing}.
\end{itemize}
This approach allows to take advantage of the full spectrum of capabilities and covers all paths in \Cref{fig:overview-compilation} except the consideration of fault-tolerant quantum computing.

\textbf{4.} A complementary approach, also employing a combination of shuttling operations and SWAP gates is \mbox{FPQA-C}~\cite{wangFPQACCompilationFramework2023}.
Based on the FPQA setup, SWAP gates are used to connect atoms, which cannot be brought close together due to the AOD constraints.
\begin{itemize}
  \item \emph{Mapping \& Scheduling:} First, the qubits are assigned to different atom arrays with one trapped in static SLM traps and the others controlled by a separate AOD each. This is done using a MAX k-cut heuristic to minimize the intra-array entangling gates, which require additional SWAP gates to connect. Inter-array gates are performed by rearranging the AOD coordinates according to \Cref{eq:shuttling_mapping} and \Cref{eq:shuttling_crossing}.
\end{itemize}
In this sense, \mbox{FPQA-C} represents a hybrid mapper for the DPQA setup, currently taking into account two-qubit gates only, resulting in $1 \rightarrow 2.A/B/D \rightarrow 3.A/B \rightarrow 3.C$.

\textbf{5.} \citet{wangQPilotFieldProgrammable2023} propose the use of so-called \emph{flying ancillas} within the FPQA setup, refered to as Q-Pilot.
Instead of bringing the gate qubits close to each other, a third ancilla qubit is shuttled between them to establish the entanglement.
In this way, trapping only the ancilla qubits in movable AODs eliminates the switching between AOD and SLM traps.
\begin{itemize}
  \item \emph{Mapping \& Scheduling:} The movements of the ancilla qubits are computed based on a heuristic taking into account \Cref{eq:shuttling_mapping} and \Cref{eq:shuttling_crossing}, with a specialized version for QAOA circuits.
\end{itemize}
While the work concentrates on mapping everything using flying ancillas, this can be considered as an alternative or complementary mapping approach using shuttling. The respective path is $1 \rightarrow 2.D \rightarrow 3./B \rightarrow 3.C$.

In addition to the capability-specific compilers, \citet{tanDepthOptimalAddressing2D2024} proposed depth-optimal addressing of single qubit gates on a two-dimensional grid using SMT solvers.
As this represents a scheduling-only problem, it can be considered in addition to any of the four capability paths, assuming the hardware supports this type of single qubit gate execution.

\subsection{Overview and Discussion}
\label{sec:overview-discussion-software}

A summarized overview of all compilers and their respective functionalities are provided in \Cref{tab:compiler_overview}, categorized based on the supported computational capabilities of \Cref{sec:comp-capabilities}.
These correspond directly to the paths in the second layer of \Cref{fig:overview-compilation} and the entailed figures of merit.
Note that this should be considered as an abstract overview at the current point in time.
The assignment to the different compiler functionality may vary depending on the exact definitions of the compilation steps and can also change in the future with the ongoing development of the tools.

However, this overview also reveals that numerous open problems still need to be addressed.
Notably, one significant challenge is the synthesis step, which involves decomposing general quantum algorithms into one- and two-qubit gates and the full set of supported multi-qubit gates.
Additionally, there is a need to explore the combination of multiple capabilities as done by for example by \mbox{FPQA-C}~\cite{wangFPQACCompilationFramework2023} and \citet{schmidHybridCircuitMapping2023}.
In this regard, in \Cref{sec:shuttling-numerics}, we perform case studies and error analysis to illustrate possible metrics to decide between multiple capabilities to perform the same operation.

In summary, while progress has been made in addressing NA-specific compilation challenges, further advancements are needed to fully exploit the platform's capabilities and simplify its utilization for both hardware experts and computer scientists.
The aim of the ``big picture'' view proposed and discussed in this section is to consider the full range of hardware capabilities instead of studying them independently of each other.
This results in compilers that take full advantage of the NAQC platform, necessary to compete with other hardware platforms.
For this aim, it is essential to evaluate promising capabilities for their usefulness and study hardware-dependent error sources to find valuable figures of merit that can be employed as cost functions during optimization.
This is done and discussed in the following \Cref{sec:numer-eval}, with complementary evaluations of the NA capabilities discussed in other recent work~\cite{wagnerBenchmarkingNeutralAtomQuantum2024,mcinroyBenchmarkingAlgorithmicPerformance2024}.

\section{Selected Case Studies}
\label{sec:numer-eval}
We now proceed to study the platform-specific capabilities of NAs in the compilation context through selected case studies. 
More concretely, we evaluate when and how much the platform-specific features and limitations impact the compilation process.
This is done by employing existing compilers from \Cref{sec:rel-work-software} and analyzing their outcome based on the approximate success probability metric outlined in \Cref{sec:errors}.
The aim is to provide estimations on error sources considering current and imminent hardware configurations.

\noindent These considerations offer insights for compiler developers, aiding in identifying predominant error sources and refining optimal compilation software.
This section does not aim at providing a final decision on the most suitable capability but to exemplarily discuss potential techniques to compare among them.
In particular, it provides information on suitable figures of merit, depending on a given hardware setup.
For hardware experts on the other side, the discussion provides a forward-looking perspective for devising future hardware designs, concentrating on the most promising capabilities within the NAQC platform.
The concepts discussed in this section can be generalized and used also for example in the context of compiling for DPQA setups.

\noindent First, a brief discussion on the error analysis based on the summary box of \Cref{sec:errors} is given, focusing on the mapping and scheduling part only. The aim is to get an easy-to-compute metric to evaluate and compare different mapping and scheduling approaches.
With this study, we can get insights about interesting figures of merit by comparing different hardware setups. 
In particular, the two most commonly used figures of merit are the questions of whether compilation passes should be optimized for gate count or circuit depth.

As a second step, we apply this analysis to numerically analyze and discuss the NA-specific capabilities, namely long-range, multi-qubit gates and shuttling.
In this case, we aim to identify the most promising capability, depending on some given hardware parameters.
This includes open questions such as the required gate fidelity or shuttling velocity to achieve an advantage for one of the two possible mapping capabilities.
Both considerations are of importance for both tool developers and hardware experts. Software tool developers get insights into how hardware parameters can affect the compilation process and change the choice of figures of merit. Concurrently, the latter can estimate the potential of given capabilities, aiding them in devising forthcoming hardware configurations.

\subsection{Error Metrics and Code Availability}
\label{sec:setup}
For the evaluations, we adopt the hardware parameters detailed in \Cref{tab:parameters}.
Following the compilation process outlined in \Cref{sec:overview-compilation}, we examine the three distinctive capabilities of long-range interactions, multi-qubit gates, and atom shuttling.
Depending on the scenario, we will vary hardware parameters, compilation constraints, and the employed compilation software to study their impact on the error metrics discussed in \Cref{sec:errors}.

\subsubsection{Fidelity Estimation}
\label{sec:error-estimations}
Due to the absence of NA-specific synthesis software, our focus will be on the mapping and scheduling stages in the subsequent sections.
We compute the final output fidelity, as introduced in \Cref{eq:success_prob}, as our performance metric.
As we will focus on the mapping and scheduling steps, the considered circuits are all fully decomposed.
Therefore, the gate errors of different mapped circuits differ only on the added SWAP gates.
When comparing different mapping approaches, it is, therefore, sufficient to focus on the fidelity reduction due to SWAP gates and idle errors instead of computing the full approximate success probability~$P$ as discussed in \Cref{sec:errors}.

\noindent For gate-based swapping, this leads to the following commutative fidelities regarding mapping and scheduling:
\begin{equation}
\label{eq:eval_gate_error}
\begin{aligned}
\mathcal{F}_\mathrm{SWAP} &= \mathcal{F}_{\mathrm{idle}} \cdot \mathcal{F}_{\mathrm{mapping}} \\
&= \exp(-\dfrac{t_{\mathrm{idle}}}{T_{\mathrm{eff}}}) (\mathcal{F}_\mathrm{CX})^{3\cdot N_{\mathrm{SWAPS}}} \, ,
\end{aligned}
\end{equation}
Here, $t_{\mathrm{idle}}$ corresponds to the idle time of the register according to \Cref{eq:idle_time}, and $N_{\mathrm{SWAPS}}$ represents the total count of introduced SWAP gates, each executed through three CX gates.
The value of $N_{\mathrm{SWAPS}}$ can be extracted directly from the compilation results.
Furthermore, we assume CX to be realizable by a combination of one single-qubit gate and a CZ gate and, therefore, add the corresponding gate durations and multiply the fidelities to get the parameters for the CX gate.
For the computation of $t_{\mathrm{idle}}$, we used Python-based post-processing scripts to correctly map and schedule the circuit gates on a given hardware. A visualization of the scheduling is available together with the scripts at Zenodo~\cite{DatasetEvaluationScripts2023}.

\noindent For shuttling-based swapping, on the other hand, we assume a process fidelity of 1 (\Cref{tab:parameters}) for the shuttling procedure, making decoherence during idle the sole source of error.
\begin{equation}
\label{eq:eval_shuttling_error}
\mathcal{F}_{\mathrm{sh}} = \exp(-\dfrac{t^{\mathrm{sh}}_{\mathrm{idle}}}{T_{\mathrm{eff}}}) \, .
\end{equation}
Here, $t^{\mathrm{sh}}_{\mathrm{idle}}$ represents the idle time of the register, incorporating the time necessary for performing the shuttling operations.

\subsubsection{Software and Code Availability}
\label{sec:softw-code-avail}
All the compilers employed in the subsequent analysis are accessible as open-source software. These tools have been installed and used in adherence to their respective documentation.
The evaluation and interpretation of data, encompassing tasks like scheduling and error computation, have been realized in Python.
The complete set of scripts employed to generate and visualize the ensuing outcomes, along with the analysis data, can be found on Zenodo~\cite{DatasetEvaluationScripts2023}.
This collection also encompasses a list of utilized software packages with their corresponding version numbers.
\vspace{1cm}

The considered quantum circuits are taken from \mbox{MQT Bench}~\cite{quetschlichMQTBenchBenchmarking2023}, a benchmarking suite containing commonly used quantum algorithms. In particular, we selected the following from the collection of benchmarks.
\begin{itemize}[itemsep=0pt,parsep=2pt]
    \item \textbf{dj:} Deutsch-Jozsa algorithm.
    \item \textbf{ghz:} Preparation circuit of the Greenberger- Horne-Zeilinger state.
    \item \textbf{graphstate:} Circuit corresponding to a \mbox{2-regular} random graph with edges representing two-qubit gates.
    \item \textbf{qft:} Quantum Fourier Transform.
    \item \textbf{twolocalrandom/two-local:} The two-local Variational Quantum Eigensolver ansatz with random parameters.
    \item \textbf{wstate:} Preparation circuit of the entangling \mbox{W-state}.
\end{itemize}

\subsection{Long-Range Interactions}
\label{sec:long-range-gate}
First, we want to study the long-range capability and the question regarding the trade-off between high connectivity and the simultaneous gate restriction. In contrast to the considerations already done in \citet{bakerExploitingLongDistanceInteractions2021a}, we focus on the question of whether the compilation pass should minimize SWAP gates or gate restriction, depending on different fidelities and coherence times.

\noindent Performing mapping with long-range interactions implies gate-based mapping incorporating a fidelity $\mathcal{F}_{\mathrm{mapping}}$ and $\mathcal{F}_{\mathrm{idle}}$, as given in \Cref{eq:eval_gate_error}.
The former arises due to the introduction of extra SWAP gates, while the latter emerges from gate durations and restrictions, resulting in idling qubits.
In the following, we aim to investigate the circumstances under which each error contribution supersedes the other, highlighting potential optimization avenues on both the hardware and software fronts.
Furthermore, we investigate, if familiar figures of merit like SWAP gate count or circuit depth are applicable for NAs.
For this aim, first, we will study different hardware configurations by varying parameters $r_{\mathrm{int}}$ and $r_{\mathrm{re}}$, followed by utilizing two compilers optimizing for different figures of merit.

\subsubsection{Interaction and Restriction Radius}

\begin{figure*}[ht]
\includegraphics[width=1\textwidth]{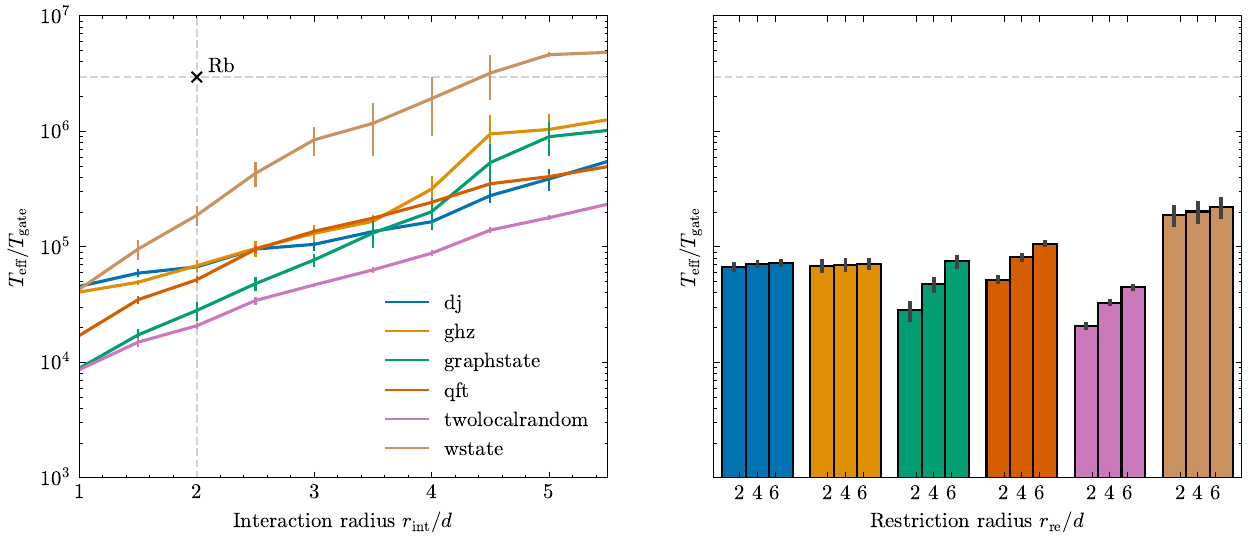}
\caption{
  \textbf{Long-range Interaction Analysis:} In the \textbf{left} graph, we computed the fidelity reduction based on SWAP gate insertion and qubit idling for diverse circuits with a dimension of $n=120$ and varying interaction radii $r_{\mathrm{int}}$.
  The displayed curve represents the effective coherence time required to equate $\mathcal{F}_\mathrm{idle}$ and $\mathcal{F}_\mathrm{mapping}$.
  This signifies that, given a circuit, when $T_{\mathrm{eff}}$ surpasses (falls below) the line, the dominant fidelity reduction is caused by SWAP gates (decoherence).
  Larger $r_{\mathrm{int}}$ diminishes the SWAP error due to augmented connectivity, necessitating larger coherence times to balance the influence of both errors.
  Owing to the prolonged coherence times of the Rubidium hardware, SWAP gates stand as the prevailing source of fidelity reduction across all circuits.
  The \textbf{right} graph follows a similar procedure, modifying the restriction factor $r_{\mathrm{re}}$ while maintaining a fixed interaction radius of $r_{\mathrm{int}}=2d$, showing different behavior depending on the structure of the circuit.
  The error bars correspond to the standard deviation averaged over 10 iterations.
}
\label{fig:long_range}
\end{figure*}

\label{sec:results-long-range}
A larger interaction radius $r_{\mathrm{int}}$ yields increased connectivity, resulting in a reduction of SWAP gates and subsequently improving $\mathcal{F}_\mathrm{mapping}$.
However, the stronger influence of long-range Rydberg interactions also imposes limitations on a greater number of nearby atoms.
As discussed in \Cref{sec:2qubitgates} this impedes gate parallelization, amplifying qubit idling, and thereby reducing $\mathcal{F}_\mathrm{idle}$.
Depending on hardware parameters, such as $T_{\mathrm{eff}}$, one of these two factors becomes the predominant source of error.
We employ the Qiskit compiler~\cite{Qiskit} to map circuits comprising $n=120$ qubits onto a square grid, followed by post-processed scheduling.
We evaluate the effects by computing the coherence time per gate duration $T_{\mathrm{eff}}/T_{\mathrm{gate}}$, where $\mathcal{F}_\mathrm{idle}$ and $\mathcal{F}_\mathrm{mapping}$ are equal. Placing hardware parameter sets in the same graph thus provides a direct guideline for future improvements.
First, we varied $r_{\mathrm{int}}$ with a fixed restriction factor $k=1$, while afterwards, we modified the restriction radius $r_{\mathrm{re}} = k r_{\mathrm{int}}$ for $k = 1,2,3$, maintaining a constant $r_{\mathrm{int}}=2d$. Each configuration is averaged over 10 runs to account for the stochastic character of the SABRE compiler.

The results are shown in \Cref{fig:long_range}. For all the algorithms considered, an increment in $r_{\mathrm{int}}$ leads to a corresponding rise in $T_{\mathrm{eff}}$ needed to achieve a balance between gate and decoherence errors.
For a given circuit, data points situated above the depicted line signify a higher $T_{\mathrm{eff}}$, indicating that the primary source of reduced fidelity is SWAP gate insertion.
Conversely, points below the line indicate a larger decoherence error.
For the Rubidium hardware configuration as detailed in \Cref{tab:parameters}, the dominance of SWAP gate errors holds true for nearly all scenarios.
Except for instances of large interaction radii, such as $r_{\mathrm{int}} \geq 4.5$ for the W-state preparation, decoherence errors begin to supersede SWAP gate errors.

On the right side of the graph, we observe how this point of error equivalence is influenced by altering the restriction radius.
Evidently, algorithms featuring a more sequential gate arrangement, like the GHZ state circuit, are less impacted by greater restrictions.
In contrast, circuits such as the Graphstate circuit or the Two-local ansatz, characterized by numerous parallel gate executions, experience more pronounced effects from an increased restriction radius.
This is due to the more sequential execution of the gates, as a parallel execution is impossible due to gate restrictions.
Subsequently, the total idle time increases and, therefore, also the decoherence error.

\noindent After varying hardware parameters and studying the influence on the output fidelity, we now want to use two compilers in the following, each optimizing for different figures of merit.

\subsubsection{Compiler comparison}
\label{sec:compiler-long-range}

\begin{figure}[ht]
\centering
\includegraphics[width=1\columnwidth]{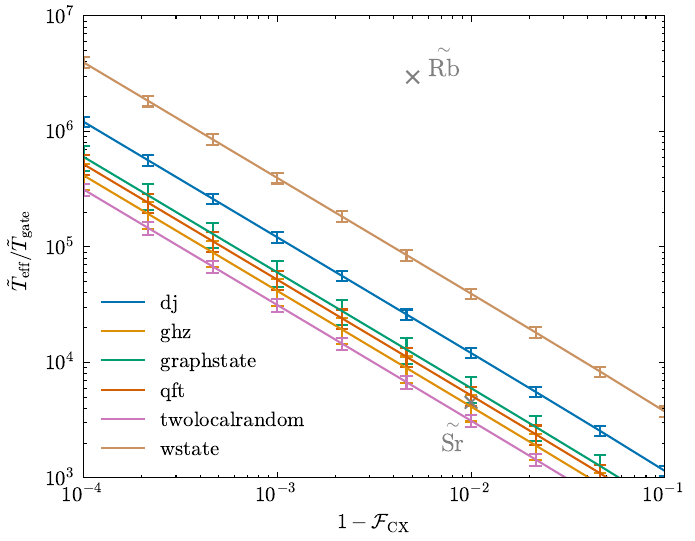}

\caption{\textbf{Comparison of Compiler Mapping Fidelity:} By utilizing the values from \Cref{tab:compiler_comp}, one can derive the gate-based fidelity reduction $\mathcal{F}_{\mathrm{SWAP}}$ for the outcomes of the compilers.
  Each circuit displays a line depicting the point at which the fidelity of both compiler outputs matches exactly while varying the CX fidelity $\mathcal{F}_{\mathrm{CX}}$.
  Given that Qiskit's compilation outcomes involve fewer SWAP gates but result in longer idle times, instances lying above the line (and therefore larger $\tilde{T}_\mathrm{eff}$) should opt for the Qiskit compiler.
  In contrast, those lying below signify lower coherence times, thus favoring Tetris to curtail qubit idling. The error bars correspond to the standard deviation over four runs of the SABRE mapping.}
\label{fig:compiler}
\end{figure}

\begin{table}[ht]
  \centering
  \scalebox{0.9}{
 \begin{tabular}{llrrrrrr}
\toprule
 &  & \multicolumn{3}{c}{ $t_\mathrm{idle} \, [\mu \mathrm{s}]$} & \multicolumn{3}{c}{$N_\mathrm{SWAP}$} \\
 &  & Qiskit & Tetris & \% & \text{    Qiskit} & Tetris & \% \\
  & $r_\mathrm{int}$ &  &  &  &  &  &  \\
\midrule
\multirow[t]{3}{*}{dj} & 1 & 187,352 & 119,887 & \bfseries{-36.0} & 314 & 500 & \bfseries{+59.0} \\
 & 2 & 113,363 & 66,703 & \bfseries{-41.2} & 152 & 372 & \bfseries{+145.1} \\
 & 3 & 72,582 & 49,099 & \bfseries{-32.4} & 84 & 209 & \bfseries{+150.0} \\
\cline{1-8}
\multirow[t]{3}{*}{ghz} & 1 & 97,847 & 74,545 & \bfseries{-23.8} & 181 & 368 & \bfseries{+102.9} \\
 & 2 & 59,990 & 52,147 & \bfseries{-13.1} & 81 & 227 & \bfseries{+179.6} \\
 & 3 & 48,855 & 37,669 & \bfseries{-22.9} & 45 & 121 & \bfseries{+167.7} \\
\cline{1-8}
\multirow[t]{3}{*}{graph} & 1 & 31,893 & 17,340 & \bfseries{-45.6} & 434 & 515 & \bfseries{+18.6} \\
 & 2 & 23,035 & 16,926 & \bfseries{-26.5} & 182 & 357 & \bfseries{+95.7} \\
 & 3 & 20,683 & 14,412 & \bfseries{-30.3} & 96 & 248 & \bfseries{+158.3} \\
\cline{1-8}
\multirow[t]{3}{*}{qft} & 1 & 1,108,553 & 813,101 & \bfseries{-26.7} & 6,117 & 8,010 & \bfseries{+31.0} \\
 & 2 & 1,177,610 & 900,108 & \bfseries{-23.6} & 2,849 & 5,576 & \bfseries{+95.7} \\
 & 3 & 1,470,671 & 979,202 & \bfseries{-33.4} & 1,479 & 3,481 & \bfseries{+135.3} \\
\cline{1-8}
\multirow[t]{3}{*}{two-local} & 1 & 2,172,426 & 1,239,620 & \bfseries{-42.9} & 29,989 & 39,882 & \bfseries{+33.0} \\
 & 2 & 2,118,934 & 1,462,344 & \bfseries{-31.0} & 14,252 & 24,102 & \bfseries{+69.1} \\
 & 3 & 2,750,812 & 2,035,308 & \bfseries{-26.0} & 7,446 & 18,188 & \bfseries{+144.3} \\
\cline{1-8}
\multirow[t]{3}{*}{wstate} & 1 & 153,827 & 94,693 & \bfseries{-38.4} & 360 & 410 & \bfseries{+13.8} \\
 & 2 & 106,314 & 72,741 & \bfseries{-31.6} & 129 & 245 & \bfseries{+89.9} \\
 & 3 & 90,755 & 68,965 & \bfseries{-24.0} & 68 & 164 & \bfseries{+139.8} \\
\cline{1-8}
\bottomrule
\end{tabular}

  }
  \caption{\textbf{Compiler Comparative Analysis:} A comparison is drawn between the Qiskit compiler~\cite{Qiskit} and the Tetris compiler~\cite{QTetris2022}, focusing on the resultant idle time $t_{\mathrm{idle}}$ and the count of inserted SWAP gates $N_{\mathrm{SWAPS}}$.
    While Qiskit prioritizes the minimization of $N_{\mathrm{SWAPS}}$, Tetris aims at enhancing parallelism, leading to a reduction of $t_{\mathrm{idle}}$.
    The final column showcases the relative variance of Tetris in comparison to Qiskit, elucidating that each compiler performs best in its respective domain.
    For Qiskit the numbers are averaged over four runs.}
  \label{tab:compiler_comp}
\end{table}
With all hardware parameters held constant, the exploration now extends to using two different compilers, each optimized for a distinct figure of merit.
In particular, we utilize the Qiskit internal SABRE heuristic~\cite{liTacklingQubitMapping2019}, which aims to minimize $N_{\mathrm{SWAPS}}$, and the Q-Tetris heuristic~\cite{QTetris2022} elaborated upon in \Cref{sec:restr-aware-comp}. All runs are repeated four times and averaged to account for the stochastic character of SABRE.
Q-Tetris orchestrates gate arrangements to amplify parallelism, hence reducing $t_{\mathrm{idle}}$.
This trend is evident in \Cref{tab:compiler_comp}, demonstrating that Q-Tetris, on average, yields a 20-30\% reduction in idle time, albeit at the expense of requiring a notably larger number of SWAP gates.
This results in up to double the SWAP gate count when compared to Qiskit.
This divergence becomes particularly pronounced for larger $r_{\mathrm{int}}$ values.
Consequently, Qiskit prioritizes minimizing $N_\mathrm{SWAPS}$ and, therefore, optimizing $\mathcal{F}_\mathrm{mapping}$, while Q-Tetris focuses on achieving optimal $t_{\mathrm{idle}}$.
Variations in CX fidelity and coherence times favor one of the two fidelities thereby selecting one of the two compilers can be favorable.

It is important to note that for this evaluation, the utilization of Q-Tetris adheres to the gate durations found in \citet{liTimingAwareQubitMapping2023}.
Specifically, it assumes that CX (SWAP) gates require 2x (6x) the execution time of a single-qubit gate.
The definition of qubit restriction aligns with that presented in \citet{bakerExploitingLongDistanceInteractions2021a}, which differs from the definition given \Cref{sec:2qubitgates}.
Consequently, we solely consider the single-qubit gate time from \Cref{tab:parameters} and compute the CX (SWAP) gate duration as the corresponding multiple.
Therefore, the hardware parameters are indicated with a tilde ($\sim$) to emphasize that they do not directly correspond to the actual hardware parameters.

As our evaluation is based on rudimentary proxy criteria for the errors in question, it is imperative to treat these findings as approximations or rough estimates.

In \Cref{fig:compiler}, the computation of the effective coherence time $\tilde{T}_{\mathrm{eff}}$, where both error components balance, is showcased.
Large coherence times increase idle fidelity $\mathcal{F}_\mathrm{idle}$, rendering the Qiskit compiler more favorable.
Hence, for data points situated above (below) the line, the Qiskit (Q-Tetris) compiler should be chosen, respectively.

When contrasting the results against the two hardware configurations listed in \Cref{tab:parameters}, it becomes evident that, for the Strontium scenario, both compilers yield comparable outcomes across most algorithms, barring the  W-state preparation and the Deutsch-Jozsa algorithm, where the Q-Tetris compiler is more suited.
This becomes clear, in particular for the W-state, considering again \Cref{tab:compiler_overview}.
Q-Tetris is able to reduce the idle time of about $\SI{60}{\milli\s}$ while adding only 50 additional SWAP gates compared to Qiskit.
For other circuits, the cost of reducing the idle time is at a larger increase in the SWAP gate number.
For these considerations, one must take into account that the fidelity difference between the two compilation outputs is based on the absolute difference in $t_\mathrm{idle}$ and $N_\mathrm{SWAPS}$, while in \Cref{tab:compiler_overview} only the relative difference is shown.
For the Rubidium hardware, on the other hand, the Qiskit compiler remains advantageous for all cases due to the extended coherence times in relation to the short gate duration.

\subsubsection{Discussion - Long-range}
\label{sec:discussion-long-range}

Examining the process of gate-based swapping reveals the emergence of two significant sources of fidelity reduction: the insertion of SWAP gates and the occurrence of decoherence during qubit idle periods.
The numerical error analysis provides insights, benefiting both hardware experts and tool developers in their efforts to improve the final fidelity of the results.
\vspace{0.9cm}

Hardware experts have the opportunity to engage in similar considerations as those demonstrated in \Cref{fig:long_range}, allowing them to analyze how hardware parameters influence the execution of specific circuits.
Moreover, it becomes feasible to estimate the potential enhancements that hardware modifications could bring about.
For instance, in \Cref{fig:long_range}, further increasing $T_{\mathrm{eff}}$ would result in only partial error improvements as the SWAP error due to small $r_{\mathrm{int}}$ overshadows the improved decoherence errors.

Compiler developers, on the other hand, will find this discourse to be of significant relevance.
While the prevailing optimization objective is often considered the reduction of inserted SWAP gates, \Cref{fig:compiler} illustrates that the key figure of merit to optimize for can substantially vary based on hardware parameters such as $T_{\mathrm{eff}}$ and $\mathcal{F}_{\mathrm{CX}}$.

This underscores the necessity for a more flexible and adaptable compiler strategy that can adjust its optimization cost function following the specific hardware configuration.

Lastly, the observation that different circuits are influenced to varying degree by the constraints imposed by gate restrictions provides advanced compilation strategies with the opportunity to choose suitable hardware configurations for each type of circuit.
Specifically, sequentially structured circuits, such as the Deutsch-Jozsa algorithm, could be scheduled for hardware setups characterized by higher $r_{\mathrm{re}}$, given that such algorithms are less sensitive to gate restriction.

\subsection{Multi-qubit Gates}
\label{sec:multi-qubit-gates-numeric}

\begin{figure}[ht]
\centering
\includegraphics[width=0.965\columnwidth]{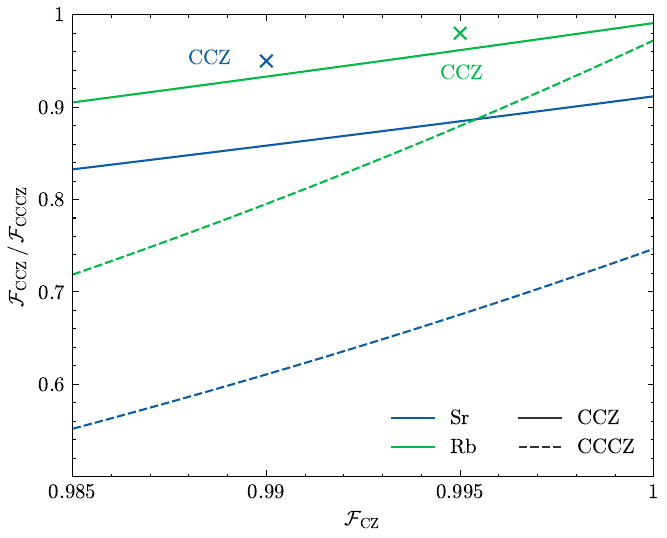}

\caption{\textbf{Required Multi-Qubit Gate Fidelity:} The lines indicate the multi-qubit gate fidelity required for the CCZ (solid) and CCCZ (dashed) gate, to be favorable to the approximate success probability (\Cref{sec:errors}) of the decomposition. The fidelity is computed for varying CZ fidelities. The respective decompositions have been found employing the Qiskit transpile function. This threshold is depicted for both hardware setups, Strontium (blue) and Rubidium (green) with the parameters taken from \Cref{tab:parameters}. For the CCZ gate, indicated by the crosses, the native implementation is preferable compared to the decomposition. The general lower success probability of the Strontium setup is due to worse single-qubit fidelity and longer gate durations.}
\label{fig:multi-qubit}
\end{figure}
The primary focus on using native multi-qubit gates is during the synthesis step.
This involves exploring whether simpler decompositions can be achieved using a broader set of native gates, including advanced gates such as the Toffoli gate.
Unfortunately, we observe currently a lack of software tools available to address this issue, and the development of such algorithms for the NAQC platform remains a mostly unresolved challenge.

Another approach is to look at circuits that already contain multi-qubit gates, for example, for reversible classical logic, and compare the native execution of the gate with the corresponding decomposition.
While multiple approaches exist for decompositions~\cite{shendeCNOTcostTOFFOLIGates2008,heDecompositionsNqubitToffoli2017}, we want to focus on direct decomposition without additional auxiliary qubits.
In this case, the multi-quit gate can be substituted one-to-one with the corresponding decomposed set of gates.
As a result, the native execution is favorable if it has a smaller error in terms of fidelity and decoherence compared to the decomposition.

To compare the native multi-qubit gates with their decomposition we employed the Qiskit transpiler function to get a decomposition of the CCZ and the CCCZ gates in terms of single-qubit gates and CZ gates. 
The decompositions are constructed of 9 single-qubit, and 6 CZ gates for the CCZ gate, and 28 single-qubit, and 20 CZ gates for the CCCZ gate. 
Subsequently, we computed the approximate success probability $P$ from \Cref{sec:errors} for the native gate and the decompositions respectively. 
\Cref{fig:multi-qubit} depicts the multi-qubit fidelity necessary depending on the CZ fidelity $\mathcal{F}_\mathrm{CZ}$. 
Additionally, the parameters from \Cref{tab:parameters} are indicated for the CCZ gate, showing that for both hardware setups, the native implementation is favorable compared to the found decomposition. This is in particular interesting for the Rubidium setup, where the CCZ gate with the corresponding fidelity has been demonstrated in experimets~\cite{evered_2023_highfidelity}. The worse performance of the Strontium setup for the same $\mathcal{F}_\mathrm{CZ}$, is mainly due to the single-qubit gates with lower fidelity and longer gate durations.

From a compilation point, this step is trivial, as the decomposition error can be computed beforehand according to \Cref{eq:success_prob_gate_only}, and depending on the outcome, the substitution can be performed or not. 
According to \Cref{fig:multi-qubit}, the native gate implementation would be favorable if its fidelity lays above the corresponding line of the decomposition.
For our considerations, the native gate execution of the CCZ gate would be preferred for both hardware setups.

\subsection{Shuttling}
\label{sec:shuttling-numerics}
This section focuses on the comparison between the two mapping capabilities, namely long-range SWAP gates or qubit shuttling. Recently, \citet{nottinghamDecomposingRoutingQuantum2023} have taken a first step toward qubit shuttling as a full replacement for SWAP gates. In the following, we want to study the possibilities of this approach in more detail. In particular, we want to study, how different hardware parameters like gate fidelity or coherence times affect the choice between the two alternatives. For this aim, we make multiple assumptions regarding the shuttling capability and analyze how the results compare to current gate-based shuttling approaches like SABRE~\cite{liTacklingQubitMapping2019}.

\noindent First, we estimate the shuttling operations as a direct replacement for the computed SWAP gates and distinguish the two cases where shuttling can be executed simultaneously to regular quantum gates or has to be scheduled in a separate shuttling layer. As a second step, we examine the possible advantage of reconfiguring all qubits between different layers. Finally, we perform an error analysis on the shuttling velocity required for shuttling to be directly superior to a similar SWAP gate.

First, we make some assumptions about the shuttling error and the possibilities of parallelization for shuttling to simplify the analysis.  For the gate-based shuttling approach, one has to consider errors stemming from imperfect CX gates as well as decoherence errors originating from qubit idling, as outlined in \Cref{eq:eval_gate_error}.
In the case of perfect shuttling, only decoherence errors remain, as per \Cref{eq:eval_shuttling_error}.
Our interest here is in investigating a simplified model, wherein each SWAP operation between physical qubits $Q_{i}$ and $Q_{j}$ is replaced by a corresponding shuttling operation characterized by a duration approximated by:

\begin{equation}
\label{eq:eval_shuttling_duration}
t^{\mathrm{sh}}(\mathrm{SWAP}(Q_{i},Q_{j})) = 2 \left(t_{\mathrm{trap}} + \dfrac{d(Q_{i},Q_{j})}{v_{\mathrm{s}}}\right) .
\end{equation}

In this equation, $t_{\mathrm{trap}}$ denotes the time required for transitioning between traps from SLM to AOD to initiate the shuttling and then back from AOD to SLM, $d$ is the physical separation distance between the two qubits, and $v_{\mathrm{s}}$ stands for the maximal shuttling velocity. The factor of two takes into account the fact that the two shuttling operations required to swap the qubits have to be executed sequentially due to the non-cross shuttling constraint of \Cref{eq:shuttling_crossing}. An illustration of this assumption is given in \Cref{fig:shuttling_substitution}.

Considering the first hardware setup of Strontium atoms in \Cref{tab:parameters}, we can estimate the shuttling time for a nearest neighbor SWAP operation as 
$$
t^\mathrm{sh} = 2 \left( 2\cdot \SI{20}{\micro\s} + \dfrac{\SI{3}{\micro\m}}{\SI{0.025}{\micro\m\per\micro\s}} \right) = \SI{160}{\micro\s} \, .
$$
This is significantly faster than a corresponding gate-based SWAP consisting of three CX gates: 
$$
t_\mathrm{SWAP}\approx 3 \cdot (\SI{200}{\micro\s} + \SI{0.2}{\micro\s}) \approx \SI{600}{\micro\s} \, .
$$
As a result, for the Strontium setup, the shuttling SWAP operation is always preferable, as it is both faster and has a higher fidelity. Therefore, we will focus mainly on the second hardware setup based on Rubidium in the following.

\subsubsection{Shuttling-based Mapping}
\begin{figure*}[ht]
\centering
\includegraphics[width=0.7\textwidth]{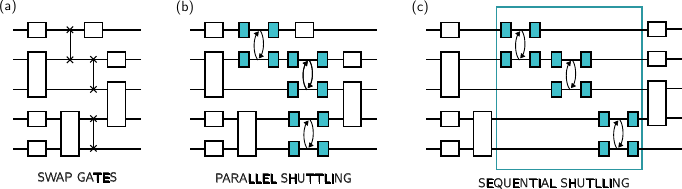}
\caption{\textbf{Shuttling Scenarios:} \textbf{(a)} Mapping output using a gate-based approach. \textbf{(b) Parallel shuttling:} The SWAP gates are replaced by shuttling operations in a one-to-one fashion. 
Each shuttling operation consists of initial trap-switch operations, the atom shuttling, and the final trap-switch operations. 
In this scenario, shuttling operations can be performed in parallel to each other and also to regular gate operations, requiring local qubit control at the hardware level. 
\textbf{(c) Sequential shuttling:} The same shuttling operations are performed but in a sequential order. 
Furthermore, regular gates and shuttling are executed in a non-parallel fashion, resulting in shuttling layers indicated by the cyan box.}
\label{fig:shuttling_strategies}
\end{figure*}

\label{sec:shuttl-based-mapp}
To draw a comparative analysis between the gate-based and shuttling-based mapping strategies, we intend to once again introduce variations in different hardware parameters favorable to one of the two cases.
Specifically, we aim to delve into the influence of two parameters: the CX fidelity, denoted as $\mathcal{F}_{\mathrm{CX}}$, which substantially impacts the precision of gate-based shuttling, and, on the other side, the effective coherence time $T_{\mathrm{eff}}$, corresponding to the decoherence errors occurring during shuttling.
For the following evaluations, we always use a fixed circuit size of $n=80$ qubits.
This analysis maintains the assumption of equal values for $r_{\mathrm{int}}$ and $r_{\mathrm{re}}$, fixed at $d$.

\noindent To perform our studies, we undertake circuit mapping using the SABRE compiler. Subsequently, following the considerations provided earlier, we engage in a post-processing step. 
This step encompasses the consolidation of consecutive SWAP gates, thereby generating the requisite shuttling operations, each with its corresponding shuttling distance.
The duration for each shuttling operation is subsequently calculated following the formula delineated in \Cref{eq:eval_shuttling_duration}.

Within the realm of scheduling, we consider two distinct scenarios, both depicted in \Cref{fig:shuttling_strategies} in comparison to a SWAP gate-based mapping.

In the first scenario, labeled as \emph{parallel shuttling}, we presuppose the availability of a sufficiently high number of independent AODs.
This premise allows for the concurrent execution of all shuttling operations that pertain to a non-overlapping set of qubits.
Moreover, we assume that both gate and shuttling operations can be executed simultaneously, requiring local qubit control on the hardware level.

Conversely, in the second scenario, referred to as \emph{sequential shuttling}, we operate under the constraint of possessing only a solitary AOD.
Furthermore, the simultaneous execution of gate and shuttling operations is restricted, meaning shuttling and gate operations have to be executed in an alternating fashion, similar to the DPQA setting of \Cref{fig:dpqa}.
In this configuration, the circuit operation involves executing all gates feasible to the current qubit layout, succeeded by another layer encompassing the sequentially executed shuttling operations, resulting in \emph{shuttling layers}, discussed in more detail in \Cref{sec:shuttl-layers}.

It is important to note that the differentiation between these two approaches resides solely within the scheduling component, with the actual shuttling operations remaining unaltered.

\begin{figure}[t]
\centering
\includegraphics[width=1\columnwidth]{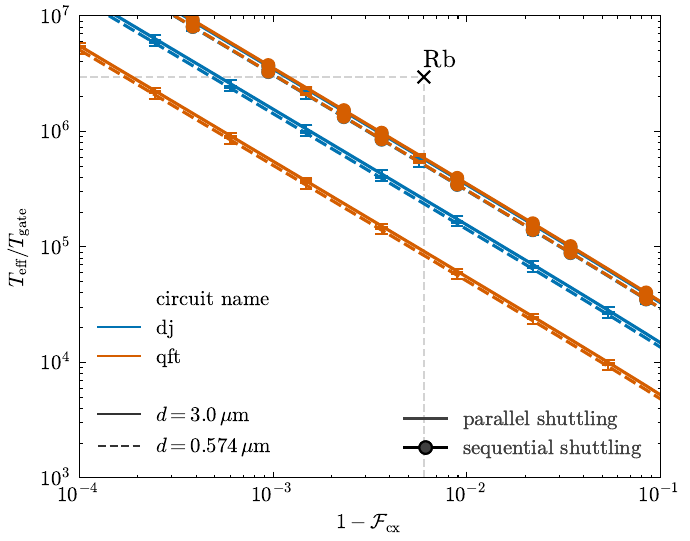}
\caption{\textbf{Error Analysis for Shuttling:} A comparative evaluation of SWAP gate- and shuttling-based Mapping is presented.
  The coherence time at which the gate-based fidelity $\mathcal{F}_\mathrm{SWAP}$ equates to the shuttling-based $\mathcal{F}_\mathrm{sh}$ is showcased by aline, corresponding to two different circuits while varying the CX fidelities.
  The solid lines depict scenarios where shuttling and gate operations can be executed concurrently, while the line indicated with a bullet represents the case where shuttling and gate executions are performed alternately. 
  The lines of sequential shuttling lay on top of each other, as due to the sequential execution of all shuttling operations the difference between circuit structures is reduced.
  Additionally, two different inter-qubit distances $d$ are distinguished, indicated by solid and dashed lines. The error bars indicate the standard deviation averaged over 20 iterations.}
\label{fig:shuttling}
\end{figure}

Depicted in \Cref{fig:shuttling} is the boundary that delineates an equilibrium point wherein gate-based and shuttling-based methodologies yield precisely the same total fidelity.
This assertion aligns with earlier discussions, signifying that data points situated above the depicted lines exhibit elevated coherence times and, therefore, favor a shuttling-based approach.
Notably, higher CX fidelities necessitate elevated values of $T_{\mathrm{eff}}$ to render shuttling-based mapping competitive.

Within the context of parallel shuttling, as denoted by the continuous lines, a difference between different circuits is visible.
Circuits inherently characterized by a sequential architecture, exemplified by cases like the Deutsch Jozsa algorithm (blue), diverge from circuits with more parallel structures, such as the Quantum Fourier Transform(red), which benefits from the concurrent execution of shuttling operations.
Conversely, when confined to a sole AOD, as indicated by the circled lines, disparities between these categories tend to attenuate, as in this case, all operations are executed sequentially.
In addition, we varied the inter-qubit distance $d$ between the value $d=\SI{3}{\micro\m}$ as given in \Cref{tab:parameters} and a prospective value of $d=\SI{0.574}{\micro\m}$.
For smaller $d$ (dashed line) lower $T_\mathrm{eff}$ are required, as the shuttling takes less time.
Nevertheless, the difference is small, as SABRE accounts only for nearest neighbor SWAPS and in this case, the shuttling duration is almost neglectable compared to the trap switching time of $t_\mathrm{trap} = \SI{40}{\micro\m}$.

\noindent Regarding the Rubidium hardware configuration, the large $T_{\mathrm{eff}}$ favors shuttling-based swapping in all cases compared to the error-prone gate-based swapping. 
CX fidelities of \mbox{$\mathcal{F}_{\mathrm{cx}} \approx 0.999$} would be necessary to make gate-based swapping a comparable alternative again.

Nonetheless, as this approach uses the SC-specific SABRE algorithm to find necessary SWAP gates, it does not take full advantage of the shuttling capability, which can also perform non-local and non-trivial SWAP operations. We will take account of this shortcoming in the following section.

\subsubsection{Shuttling Layers}
\label{sec:shuttl-layers}
Now, we want to consider the second shuttling scenario of sequential, layer-based shuttling.
This separation between gate execution and shuttling facilitates the examination of their respective impacts on the overall idle time, denoted as $t_{\mathrm{idle}}$, and thereby, the resultant decoherence error.

\begin{figure}[t]
\centering
\includegraphics[width=1\columnwidth]{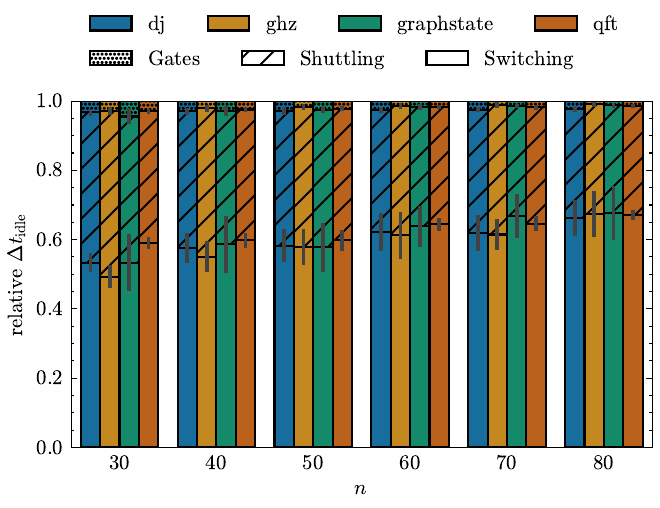}

\caption{\textbf{Comparative Analysis of Idle Times:} The correlation between shuttling and gate execution times in layer-based shuttling is depicted as a ratio.
  This ratio is influenced by distinct circuit types and increasing circuit sizes ($n$).
  The analysis takes into account the Rubidium hardware configuration for both shuttling and gate timings.
  Remarkably, over 90\% of the overall circuit execution time is attributed to shuttling and trap-switching operations, with this proportion rising as the number of qubits increases. 
  The large contribution of gate-switch is caused by SABRE preferring multiple short-range SWAP gates, each requiring the corresponding trap switch duration $t_\mathrm{trap}$.
  The error bars indicate the standard deviation over 50 iterations.}
\label{fig:idle_times}
\end{figure}

As depicted in \Cref{fig:idle_times}, the graph portrays the proportion of idling attributed to gate-related factors versus idling linked to shuttling operations across different circuits and varying numbers of qubits.
For all instances considered, the contribution from shuttling operations accounts for more than 90\% of the cumulative idle time, in particular for larger qubit numbers.
Furthermore, it shows that for our assumptions more than half of the total idle time is due to trap switching.
Here, it should be noted that the long trap switching times are due to SABRE, with its SC background, preferring multiple short-distance SWAPs compared to a single long-range operation.
So the results can be expected to be improved by employing a shuttling-specific compiler.

This method, involving the utilization of SABRE followed by the subsequent substitution of resultant SWAPs with shuttling operations, does, therefore, not use the full shuttling potential.
SABRE consistently opts for the shortest feasible swapping path to minimize the count of SWAP gates.
Conversely, for shuttling, it may be strategically advantageous to transport a qubit across a greater distance, potentially improving the future requirement for swapping.
To address this prospect, we modify the mapping process, whereby SABRE is initially employed to ascertain a mapping configuration, enabling the execution of as many gates as feasible.
Subsequently, a wholly fresh initial mapping is determined for the remaining circuit.
Then again, all feasible gates are executed, and the process of finding a new mapping for the next layer is iterated until the complete circuit is processed.

\begin{figure}[t]
\centering
\includegraphics[width=1\columnwidth]{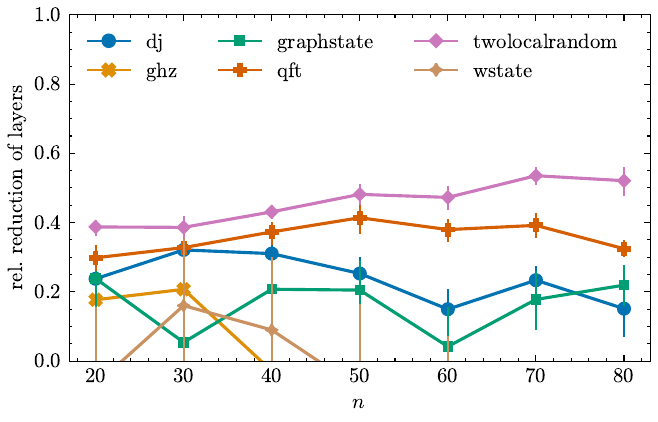}

\caption{\textbf{Layer Reduction via Full Reconfiguration:} The lines present the ratio by which the number of layers is reduced when the mapping algorithm is granted complete reconfigurability for each layer.
This demonstrates the potential improvement by utilizing long-range shuttling compared to nearest neighbor SWAPs.
The error bars indicate the standard deviation over 50 iterations.}
\label{fig:layers_shuttling}
\end{figure}

As depicted in \Cref{fig:layers_shuttling}, the illustration showcases the proportion by which the cumulative count of layers can be diminished by employing this comprehensive qubit reconfiguration for each layer.
Depending on the specific circuit and the count of qubits, this approach can yield reductions of up to 50\% in the number of shuttling layers.
While it is generally expected that the required inter-layer shuttling becomes more complicated with this technique, it still underscores a promising benefit of utilizing the full reconfiguration capabilities of NAQC platforms.

Until now, we assumed fixed shuttling hardware parameters and only modified $T_{\mathrm{eff}}$. But also other parameters such as the trap switching time or the actual shuttling velocity $v_{\mathrm{sh}}$ affect the shuttling time of \Cref{eq:eval_shuttling_duration} and therefore the shuttling error. In the following, we want to study these parameters by the question, how fast must the shuttling be to be favorable compared to a corresponding gate-based SWAP operation?

\subsubsection{Shuttling Velocity}
\label{sec:shuttling-velocity}
The shuttling duration, as expressed in \Cref{eq:eval_shuttling_duration}, is determined by two components: the time $t_{\mathrm{trap}}$ required to transition between trap types, i.e.~SLM to AOD and vice versa, and the shuttling velocity $v_{\mathrm{sh}}$.
When considering a specific CX fidelity $\mathcal{F}_{\mathrm{CX}}$, the duration of shuttling must be less than a certain critical value for shuttling-based SWAP to exhibit improved error performance compared to the equivalent SWAP gate.

In the context of the Rubidium hardware, even though the time taken for shuttling-based SWAP surpasses that of three consecutive CX gates, the large $T_{\mathrm{eff}}$ mitigates decoherence errors, particularly when only two qubits are involved.
However, with larger values of $n$, which also implies more qubits idling during the shuttling SWAP, a crossover exists where less accurate yet faster CX gates become the preferred choice.
This is of particular interest for SWAP operations that can not be parallelized with other gate operations, occurring often, particularly for sequentially structured algorithms.

\begin{figure}[t]
\centering
\includegraphics[width=1\columnwidth]{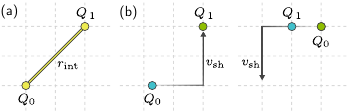}
\caption{\textbf{Shuttling as Direct SWAP Substitution:} Illustration of the estimations taken in \Cref{fig:shuttling_speed}. \textbf{(a)}~$\mathrm{SWAP}(Q_0,Q_1)$ executed as a long-range interaction over distance $r_\mathrm{int}$. \textbf{(b)}~Illustration of a two-step shuttling process for the same SWAP operation, using shuttling. Due to \Cref{eq:shuttling_crossing} these operations can not be executed exactly as shown, but are used as a first approximation of the actually required operations.}
\label{fig:shuttling_substitution}
\end{figure}

An illustration of this direct substitution of gate-based SWAP operations by the corresponding shuttling operations is shown in \Cref{fig:shuttling_substitution}, where the $\mathrm{SWAP}(Q_0,Q_1)$ gate over a distance $r_\mathrm{int}$ is substituted by two shuttling operations.

\begin{figure}[t]
\centering
\includegraphics[width=1\columnwidth]{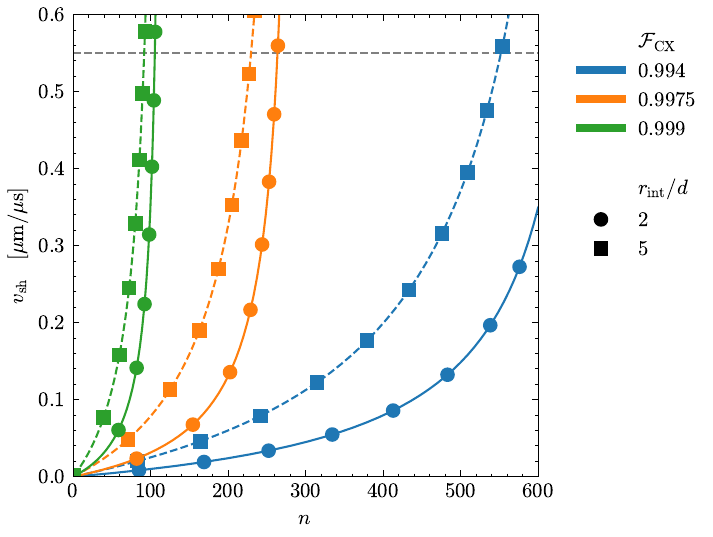}
\caption{\textbf{Shuttling Velocities:} Depicted here is the minimum necessary shuttling velocity to attain an error equivalent to that induced by gate-based swapping for a single swap pair for the Rubidium hardware setting. The x-axis indicates the number of qubits $n$ which must idle during the shuttling operation.
For more idle qubits, and therefore increased idle errors, the shuttling speed must also increase to keep the overall shuttling error the same. 
The steep increase of $v_\mathrm{sh}$ is due to the fact that the shuttling time has a lower limit by the trap switching time $t_\mathrm{trap}$.
For an increased interaction radius $r_\mathrm{int}$, the shuttling has to be performed faster due to the longer shuttling distance.
While shuttling is a suitable substitute for SWAP gates with a current fidelity of 0.995, even for high idle qubit numbers of 500, an increase in $\mathcal{F}_\mathrm{CX}$ favors the gate-based swapping again.}
\label{fig:shuttling_speed}
\end{figure}

\Cref{fig:shuttling_speed} showcases the required shuttling velocity ($v_{\mathrm{sh}}$) to achieve parity of $\mathcal{F}_\mathrm{SWAP}$ and $\mathcal{F}_\mathrm{sh}$ for a single SWAP operation, depending upon the count of idle qubits.
Points situated above the lines denote scenarios where the implementation of shuttling for SWAP gates results in improved shuttling fidelity compared to the gate-based alternative.
Within the context of the Rubidium scenario ($\mathcal{F}_{\mathrm{CX}}=0.995$ and $r_{\mathrm{int}}=2d$, shown in blue), the critical number of idle qubits is determined by the intersection with the maximal shuttling velocity ($v_{\mathrm{sh}}= \SI{0.55}{\micro\m\per\micro\s}$), occurring at approximately 550 qubits.
As examples, higher CX fidelities and interaction radii are depicted, both reducing further the maximal number of allowed idle qubits.
While this consideration does not take into account parallel shuttling, it still shows the possibility of preferring SWAP gates compared to shuttling for near-term hardware with qubit numbers in the range of multiple hundreds.
The advantages of a shuttling-based approach for shuttling over longer distances and in parallel will be subject to future work, requiring a fully working shuttling compiler.

\subsubsection{Discussion - Shuttling}
\label{sec:discussion-shuttling}
The approximations conducted in this section underscore the substantial promise inherent to atom shuttling as a prospective substitute for gate-based mapping.
This particularly holds true for scenarios characterized by large $T_{\mathrm{eff}}$ and lower $\mathcal{F}_{\mathrm{CX}}$.
These insights can be beneficial for hardware experts striving to enhance available hardware capabilities.
Considerations like the layer-based complete reconfiguration, as demonstrated in \Cref{fig:layers_shuttling}, underscore the latent advantages of employing mapping strategies specifically tailored to shuttling.

For compiler developers, this introduces a range of implications.
Initially, compilation strategies must ascertain the availability of shuttling operations, and subsequently devise mapping strategies that effectively leverage the capacity to perform qubit swaps over significant distances in parallel.
Recently, pioneering work in this direction has been made~\cite{OLSQDPQACompiler2023,nottinghamDecomposingRoutingQuantum2023}.
Our calculations indicate, furthermore, that for the Rubidium hardware setting with a single AOD, the shuttling and trap switch process constitutes the central part of qubit idling.
Additionally, we show that using the capability of NAs to completely rearrange all qubits after each gate layer can reduce the total number of shuttling layers by up to 50\%.
In this way, we delineate the prospective advantage of adopting a fully reconfigurable mapping strategy instead of the conventional nearest-neighbor methodology.

Furthermore, the evaluations presented in \Cref{fig:shuttling_speed} offer indications that hardware parameters such as $T_{\mathrm{eff}}$, $\mathcal{F}_{\mathrm{CX}}$, $r_{\mathrm{int}}$, and the number of idling qubits $n$ impact the decision of whether a SWAP operation should be executed using the slower yet accurate shuttling method or faster gate operations, although error-prone.
This accentuates the need for a hybrid compiler strategy capable of making informed choices based on the underlying hardware parameters, guiding the execution of remapping through either of the two available capabilities.
For decision-making, a similar approach as proposed in this work can be employed to find the optimal compilation process, given the available hardware.

\subsection{Discussion}
\label{sec:discussion}
In this section, we performed multiple selected case studies using different hardware parameters to study the compilation output error for the available NA capabilities.

\noindent For the possibility of long-range SWAP gates, we focused on comparing fidelity reduction due to SWAP gates and decoherence on the other hand.
Our considerations indicate that for the Rubidium setup, the SWAP gate errors dominate for all considered circuits.
In this scenario, high connectivity with $r_{\mathrm{int}} \geq 5d$ would be necessary to get comparable error contributions from decoherence errors.
This also implies that SWAP gate minimization can be considered a suitable figure of merit, making common SC compilers such as SABRE and their techniques interesting candidates.
For the first hardware setup, based on Strontium atoms, on the other hand, both errors contribute equally, requiring simultaneous optimization of the SWAP gate and idle-time minimization.
This implies the need for hardware adaptive compilers that can optimize different figures of merit depending on given hardware parameters and, therefore, the primary source of errors.

\noindent For multi-qubit gates, the task of synthesis remains an open question. 
In contrast, a simple error estimation can be performed for circuits already containing multi-qubit gates to decide if the gate should be executed natively or decomposed.

\noindent For the shuttling capability, we make multiple simplifying assumptions to illustrate the promising potential as an alternative to regular gate-based swapping.
We compare the two swapping techniques concerning different hardware parameters such as coherence times, CX gate fidelity, shuttling velocity, and qubit number.
As a result, shuttling outperforms gate-based swapping for both considered hardware setups.
Nevertheless, we also indicate possible situations regarding high CX fidelity or a large number of idle qubits, where the error-prone but faster SWAP gates become interesting again.
This would imply the need for a hybrid compilation process, where the mapping pass decides dynamically if gate-based swapping or shuttling-based qubit rearrangement is favorable.

\noindent In summary, the evaluations performed in this section illustrate the use of the capabilities discussed previously and give multiple insights on how different hardware parameters affect the compilation output.
This is helpful for tool developers to build hardware-aware compilation software based on optimization techniques based on valuable figures of merit and, this way, facilitates the development of NA-specific compilers.
At the same time, the results can also give hardware experts insight into devising future hardware, prioritizing hardware attributes that yield the most likely output improvement.
As an advantage, this allows for the effective co-design of hardware setups and compilation software, necessary to explore the full capabilities of the NAQC platform.

\section{Summary and Outlook}
In this work, we studied the overall compiler development for the Neutral Atom Quantum Computing platform and provided important groundwork to promote further collaboration between hardware experts and computer scientists.

Initially, we expounded upon the foundational physical aspects integral to realizing quantum processors utilizing neutral atoms, explicitly emphasizing the distinct computational capabilities intrinsic to the platform.
We provided abstraction layers that cater to the design automation community and software tool developers, facilitating their comprehension and abstraction of the physical processes.
Subsequently, we delved into the structural organization of this spectrum of compilation strategies and explored figures of merit for assessing the quality of the resultant compilation outcomes.
Furthermore, we furnished an overarching view of the currently available software tools and compilers, contextualizing their roles within the previously discussed compilation overview.
Lastly, we performed multiple selective case studies and fidelity analyses to investigate the implications of different hardware parameters on the final compilation outcome.
In particular, we evaluate and compare the different capabilities and match the results to two possible hardware setups, giving insights to both hardware experts and tool developers.
The results underscore the imperative to develop hybrid and \mbox{hardware-aware} compilation software capable of effectively addressing the broad spectrum of capabilities offered by the neutral atom platform.

We posit that this comprehensive overview can effectively contribute to the development of top-tier compilers and design automation tools, enabling the neutral atom platform to catch up with comparable solutions available for other hardware platforms.
In particular, the next steps entail the development of new compilation software, based on the information gained within this work regarding possible hardware capabilities and useful figures of merit.

Furthermore, neutral atoms have been shown to be promising candidates for fault-tolerant quantum computing~\cite{xuConstantOverheadFaultTolerantQuantum2023,bluvsteinLogicalQuantumProcessor2023} due to their extended range of capabilities. 
This includes high gate fidelities, combined with e.g. the ability to perform native multi-qubit gates, mid-circuit measurements, and qubit shuttling.
In particular, the qubit shuttling allows beyond \mbox{nearest-neighbor} connectivity without breaking fault tolerance due to error propagation such as SWAP gates, and the simultaneous control of large qubit patches using parallel AOD movements.
The fundamental basics discussed in this work also pave the way toward fault-tolerant compilation, which will add another layer of complexity to the compilation chain, and finding suitable automation techniques is still an open question.

\begin{acknowledgments}

All authors acknowledge funding from the Munich Quantum Valley initiative (K3, K5, K8), which is supported by the Bavarian state government with funds from the Hightech Agenda Bayern Plus.
L.S. and R.W. acknowledge funding from the European Research Council (ERC) under the European Union's Horizon 2020 research and innovation program (Grant Agreement No. 101001318).
D.F.L, and M.M. acknowledge support by the BMBF project IQuAn.
M.R. and M.M. also acknowledge funding by the Deutsche Forschungsgemeinschaft under Germany's Excellence Strategy ``Cluster of Excellence Matter and Light for Quantum Computing (ML4Q)'' EXC-2004/1-390534769.
J.Z. and S.B. acknowledge funding by the Max Planck Society (MPG), the Deutsche Forschungsgemeinschaft (DFG, German Research Foundation) under Germany’s Excellence Strategy EXC-2111-39081486. J.Z. acknowledges support from the BMBF through the program ``Quantum technologies - from basic research to market'' (SNAQC, Grant No. 13N16265).
D.F.L.,  M.M., J.Z. and S.B. additionally acknowledge support by the BMBF project MUNIQC-ATOMS (Grant No. 13N16070).
S.B. and J.Z. are co-founders and shareholders of planqc.
\vspace{4cm}
\end{acknowledgments}

\bibliography{refs}

\end{document}